\documentclass[journal,10pt]{IEEEtran}
\usepackage{amsfonts}
\IEEEoverridecommandlockouts

\usepackage{epsfig}
\usepackage{graphicx}
\usepackage{psfig}
\usepackage{subfigure}
\usepackage{epsf}
\usepackage{epstopdf}
\usepackage{amssymb}
\usepackage[cmex10]{amsmath}
\usepackage{booktabs}
\usepackage{fancyhdr}

\usepackage{algorithm}
\usepackage{algorithmic}
\usepackage{epstopdf,cite,color}
\usepackage{amsthm}
\newcommand{\bc}{\begin{center}}
	\newcommand{\ec}{\end{center}}
\newcommand{\be}{\begin{equation}}
\newcommand{\ee}{\end{equation}}
\newcommand{\bea}{\begin{eqnarray}}
\newcommand{\eea}{\end{eqnarray}}

\renewcommand\IEEEkeywordsname{Index Terms}

\begin{document}
	
	\title{ Energy Efficiency and Delay Tradeoff in an MEC-Enabled Mobile IoT Network}
	\author{Han~Hu,~\IEEEmembership{Member,~IEEE},
		Weiwei~Song,~Qun~Wang,~\IEEEmembership{Student Member,~IEEE}, Rose~Qingyang~Hu,~\IEEEmembership{Fellow,~IEEE},~and Hongbo~Zhu
		
		\thanks{H.~Hu, W.~Song, and H.~Zhu are with the Jiangsu Key Laboratory of
			Wireless Communications, Nanjing University of Posts and Telecommunications, Nanjing 210003, China, and also with the Engineering Research Center of Health Service System Based on Ubiquitous Wireless Networks, Ministry of Education, Nanjing University of Posts and Telecommunications, Nanjing 210003, China (e-mail:han\_h@njupt.edu.cn, 1018010111@njupt.edu.cn, hbz@njupt.edu.cn)
		
             Q.~Wang~and~R.~Q.~Hu are with the Department of Electrical and Computer Engineering, Utah State University, Logan, UT, USA. (e-mail: claudqunwang@ieee.org,  rose.hu@usu.edu.)	}
	}
	\maketitle
	
	\IEEEpeerreviewmaketitle
	\begin{abstract}
 Mobile Edge Computing (MEC) has recently emerged as a promising technology in the 5G era. It is deemed an effective paradigm to support computation intensive and delay critical applications even at energy-constrained and computation-limited Internet of Things (IoT) devices. To effectively exploit the performance benefits enabled by MEC, it is imperative  to jointly allocate radio and computational resources  by considering  non-stationary  computation demands, user mobility, and wireless fading channels. This paper aims to study the tradeoff between energy efficiency (EE) and service delay for multi-user multi-server MEC-enabled IoT systems when provisioning offloading services in a user mobility scenario. Particularly, we formulate a stochastic optimization problem with the objective of minimizing the long-term average network EE with the constraints of the task queue stability, peak transmit power, maximum CPU-cycle frequency, and maximum user number.  To tackle the problem, we propose an online offloading and resource allocation algorithm by transforming the original problem into several individual subproblems in each time slot based on Lyapunov optimization theory, which are then solved by convex decomposition and submodular methods. Theoretical analysis proves that the proposed algorithm can achieve a $[O(1/V), O(V)]$ tradeoff between EE and service delay.  Simulation results verify the theoretical analysis and demonstrate our proposed algorithm can offer much better EE-delay performance in task offloading challenges, compared to several baselines.

\end{abstract}
\IEEEpeerreviewmaketitle
\section{Introduction}
The fast advancement of Internet of Things (IoT) and 5G wireless technologies  greatly facilitate the proliferation of new IoT applications, such as augmented reality, surveillance, and e-Healthcare \cite{app}. Nevertheless, mobile IoT devices (MIDs) are normally constrained by the limited computation capacity and battery lifetime, due to the physical size limitation (e.g., wearable devices, on-device-sensors, and smartphones). Thus they may not be able to support  the computation-intensive and delay-sensitive services just by themselves.  One potential way to overcome this contradiction is to offload the computation-intensive tasks to the cloud center. However,  due to the long communication distance between the cloud and MIDs, this may not be always an effective way for the explosively growing IoT-based traffic, especially for the delay-sensitive traffic \cite{chenxu}. As such Mobile edge computing (MEC) \cite{wuwei} is envisioned as a promising paradigm to overcome these hurdles by exploiting the edge servers that are much closer to the local devices. In  wireless networks, edge nodes, such as base stations and edge routers,  can be equipped with high computational and storage capabilities to meet MIDs' requirement. Therefore, MIDs can support delay-sensitive and computation-intensive services efficiently, and their battery energies are saved for longer lifetime. 

MIDs can offload their tasks to the proximal MEC servers by radio interfaces. The latency can  include the communication time between the MID and the MEC server \cite{mao-energy}, and the queuing delay. 
The energy consumption and latency for communication at each MID are notably affected by the radio channel condition. When the channel is in a better condition, the transmission rate would be higher, then both the transmission energy and latency are reduced. When the channel is in a bad condition, the MID may wait till the channel condition turns to be good, which leads to a longer MID queuing delay. 
Besides, the requirements of service delay and energy consumption can vary in different  application scenarios. For instance, an industrial IoT (IIoT) application is very stringent on delay as a delayed response can lead to a disaster or some serious consequences \cite{indus-iot}. Same to the  autonomous driving, the requirement on energy consumption are relatively low while the latency is critical  \cite{vot}. Another example is online gaming on mobile devices, such as augmented reality. It requires low service latency and high energy efficiency \cite{game}.
It is important  to consider a good balance between energy efficiency (EE) and service delay, which has a great influence on the user's quality of experience (QoE).

Many recent studies have been focusing on  the EE-delay trade-off study in MEC systems \cite{tradeoff1,tradeoff2,blockchain, Dai-jointuserass,Tran-multiser}.
There still exist many challenges yet to be addressed. First, the wireless channel conditions are varying over time, and the future information about the channel condition as well as the stochastic task arrivals is hard to predict. Thus, how to make timely offloading decisions with all these uncertainties is an indeed challenge. Second, mobility is another important factor that can impact offloading, which further complicates the offloading design.  Due to mobility \cite{mobility}, especially in an multi-server MEC scenario, the associated MEC server would be different from time to time, depending on the position  of each MID, the available communication and computation resources between MIDs and servers in the current time slot. Third, due to all these dynamic conditions, it is not an effective choice to 
consider instantaneous performance requirements other than the long-term average system performance. 

Motivated by the challenges mentioned above, the EE-delay tradeoff for the multi-server multi-user IoT networks is investigated in this paper. We first develop a new scheme that jointly considers user mobility, dynamic wireless channel states, stochastic task arrival. We then propose a stochastic optimization algorithm to address the formulated long-term EE minimization problem while providing performance guarantees for the MIDs.

The  major contributions of the paper are summarized as follows:
\begin{itemize}
	\item Through a comprehensive consideration of the user mobility, random channel states, and stochastic task arrivals, we provide a study on the fundamental tradeoff between EE and service delay in a multi-server multi-user IoT network. 
	\item A stochastic optimization problem is formulated to optimize the long-term average EE of all the MIDs,
	subject to the constraints on the task queue stability, the maximum CPU-cycle frequency, the maximum transmit power at each MID, and maximum association number of one MEC server.
	\item We design an online offloading and resource allocation algorithm (OORAA) to efficiently solve the formulated EE-delay tradeoff problem. Specifically, we first decompose the formulated problem into several individual subproblems in each time slot by utilizing fractional programming and Lyapunov optimization. We further solve these subproblems by convex decomposition and submodular optimization methods. 
	\item Theoretical analysis proves that the proposed OORAA algorithm can balance the EE and the service delay through adjusting the control parameter $V$. Extensive simulations  demonstrate the superiority of the proposed algorithm compared to other benchmark methods.
\end{itemize}

The rest of the paper is organized as follows. Section II reviews the related works. In Section III, we describe the system model. In Section IV, the EE-delay tradeoff problem is formulated. The details of our algorithm are presented in Section V. We evaluate the performance of our algorithm through extensive experiments in Section VI, followed by the conclusion in Section VII.

\section{Related Work}

The MEC paradigm has attracted significant attention from computing and communications research communities in recent years \cite{ymaosurvey}. By offering storage and computing resources at the edge servers deployed close to the devices, the MEC system facilitates a highly accessible platform for 
low-latency content delivery and computation services \cite{jun-vot}. MEC technology has been substantially exploited in supporting  computation-intensive and latency critical IoT and 5G applications \cite{qun}. Nokia proposed an MEC architecture for UAV traffic management (UTM) to connect UAVs in \cite{UTM}. A work \cite{fuhui-uav} studied the computation rate maximization problem in a UAV-enabled MEC wireless powered system.

The performance tradeoff between the service delay and EE in an MEC system has been extensively studied \cite{mao-tradeoff, mao-multiaccess, deng-parallel,min-learning}. 
Mao \textit{et.al.} \cite{mao-tradeoff} proposed an online binary offloading algorithm by joint optimizing communication and computational resource allocation in an multi-user MEC system, which can balance the energy consumption and delay performance. Mao \textit{et.al.} \cite{mao-multiaccess}
considered two multiple access modes with TDMA and FDMA, and investigate online offloading strategies for multi-user wireless powered MEC systems, respectively. Deng  \textit{et.al.} \cite{deng-parallel} formulated a dynamic optimization problem to  minimize both the response time and packet loss, subject to energy queue stability in a green MEC framework.
Min  \textit{et.al.} \cite{min-learning} investigated MEC and offloading rate selection problem. An RL-based offloading scheme is proposed and the performance bounds in terms of energy consumption, computation latency, and utility are provided. The aforementioned works either focus on single-edge multi-user systems or single-user multi-server systems.

In recent years, several works have studied the EE-delay tradeoff of multi-server multi-user systems \cite{blockchain, Dai-jointuserass,Tran-multiser, new1, Guo-multi, He-multi,han-iot}. Feng \textit{et.al.}  \cite{blockchain} concentrated on the optimization of blockchain and MEC systems, and proposed a joint optimization algorithm to achieve the tradeoff between the energy consumption of MEC systems and the transaction delay of the blockchain system, while the sum execution time by MEC system is considered. Dai \textit{et.al.} \cite{Dai-jointuserass} formulated joint user association and computation offloading problem, and proposed an efficient computation algorithm to minimize overall energy consumption, subject to the overall delay constraint. Tran  \textit{et.al.}  \cite{Tran-multiser} studied the joint task offloading and resource allocation for multi-server MEC systems, and aimed to maximize a weighted sum of reductions in task computation delay and energy consumption. { Anajemba  \textit{et.al.} in \cite{new1} presented a cooperative offloading technique based on the Lagrangian suboptimal convergent computation offloading algorithm for multi-access MEC in a distributed IoT network. A suboptimal computational algorithm was implemented to perform task offloading.} Nevertheless, the studies \cite{blockchain,Dai-jointuserass,Tran-multiser,new1} only focus on the short-term performance of one single task. On the contrary, for some delay-tolerant systems due to the dynamic task generating process and the channel statistics, the long-term average performance should be considered. The works \cite{Guo-multi,He-multi,han-iot} studied the average long-term performance in multi-server MEC networks. Guo \textit{et.al.} \cite{Guo-multi} formulated a dynamic binary offloading problem to minimize the average task execution delay subject to the average energy consumption in time-slotted multi-server MEC system by taking advantage of Lypapunov and multi-armed bandit theory. 
He \textit{et.al.}  \cite{He-multi}  designed a novel binary offloading scheme that can preserve user privacy while enhance the energy cost under the constraint of delay to improve user experience. Our previous work \cite{han-iot} proposed an online computation offloading scheme for the multi-server MEC systems with energy harvesting devices based on Lyapunov optimization and semi-definite programming.
However, most of them did not consider the partial offloading model  \cite{Guo-multi,He-multi}, and ignore energy consumption per bit \cite{Guo-multi,He-multi,han-iot}.

User mobility is also a critical factor in the MEC system design \cite{follow, zhan,tan,mobility-cache,mobilitysun} in recent years. The authors in \cite{follow} designed a mobility-aware dynamic service placement scheme to minimize the long-term service latency with a long-term time-averaged handover cost budget. In \cite{zhan}, the authors defined a novel user experience utility as the improvement of the user experience over local execution, proposed a heuristic mobility-aware binary offloading scheme to achieve an optimal system-wide user utility. In \cite{tan}, the authors proposed the edge caching and processing framework for connected vehicle and designed a deep Q-learning resource allocation algorithm to minimize the system cost, subject to the limited and dynamic storage and computational resources at the vehicles. The authors in \cite{mobility-cache} modeled the interactions between caching vehicles and mobile users as a two-dimensional Markov process, and proposed an online vehicular caching algorithm toward the minimization of EE. The authors in \cite{mobilitysun} focused on a typical user, and developed a user-centric energy-aware mobility management strategy to determine which BS to associate and when to perform handover. None of these works explored the online offloading and resource scheduling problem. 

It is important to study the optimal policies of mobility-aware offloading in multi-server multi-user IoT networks. Many existing studies focus only on a single MEC server and quasi-static scenarios while the challenges brought in by dynamic task arrival and wireless channel environment are not addressed adequately. Furthermore, the user-association policy is a critical issue in the mobility-aware computation offloading design. Different from the existing methods, the proposed scheme  not only balances EE and service delay in a multi-server multi-user IoT network but also takes the dynamic environment into account. This work  handles the mobility and stochastic optimization towards performance tradeoff in mobile edge computing.

\section{System Model}
\label{systemmodel}
\begin{figure}[h]
	\vspace{-0.3cm}
	\setlength{\abovecaptionskip}{-0.2cm} 
	\setlength{\belowcaptionskip}{-1cm}
	\centering
	\includegraphics[width=3.2in]{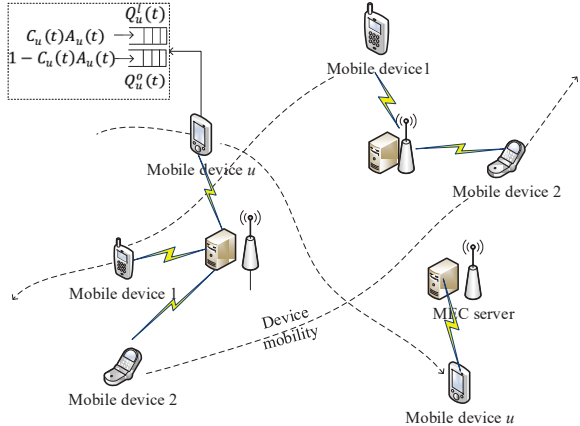}
	\caption{System model.}\label{sym}
\end{figure}

As shown in Fig. 1, we consider an MEC-enabled IoT system consisting of $M$  Small Base Stations (SBSs), denoted as $\mathcal{M} = \{ 1,2,...,M\}$, and  $U$ MIDs, denoted as  $\mathcal{U} = \{ 1,2,...,U\}$.
Each SBS is endowed with an edge server to provide both radio access and computation services to the resource-constrained MIDs. By offloading full or partial computation tasks to a MEC server, the MIDs can save battery life and also reduce computation latency.

We focus on a multi-user mobile scenario where the MIDs move randomly whereas the MEC servers keep static. In order to capture the stochastic movement, we take the widely used {\color{red}} basic model, i.e., random walk model, to generate the trajectory of MIDs as in the work \cite{mobilitysun}. 
The time is assumed to be discretized into time slots, indexed by $t \in {\rm \mathcal{T}} = \{ 1,2,...,T\}$ with a time slot length $\tau$. At the beginning of each time slot, MIDs will decide which MEC server to associate. Thus, we use a binary indicator  $x_{um}(t)$ to denote the user association decision. If MID $u \in \mathcal{U}$ decides to offload tasks to MEC server $m$, then  $x_{um}(t)=1$; otherwise $x_{um}(t)=0$.  Note that at a given time slot, each MID can only associate with at most one MEC server and each MEC server can concurrently serve at most $N_{max}$ MIDs. Thus, we have the following constraints for $x_{um}(t)$.

\be
	\setlength{\abovedisplayskip}{3pt}
\setlength{\belowdisplayskip}{3pt}
 \sum_{m \in \mathcal{M}} {x_{um}(t) \le 1,\forall u \in \mathcal{U},t \in \mathcal{T}}, 
\ee
\be
		\setlength{\abovedisplayskip}{3pt}
	\setlength{\belowdisplayskip}{3pt}
x_{um}(t) \in \{ 0,1\} ,\forall m \in \mathcal{M},u \in \mathcal{U},t  \in \mathcal{T}.
\ee
\be
		\setlength{\abovedisplayskip}{3pt}
	\setlength{\belowdisplayskip}{3pt}
\sum\limits_{u = 1}^U {x_{um}}(t) \le N_{max}, \forall  m\in \mathcal{M},t \in \mathcal{T}.
\ee

Let $A_u(t)$ (bits) denote the computation task arrived at MID $u$.  Without loss of generality, we assume that $A_u(t)$ is bounded by $[A_{\min},A_{\max}]$ with $\mathbb{E}[A_u(t)]=\lambda_u,u \in \mathcal{U}$.  We consider the partial offloading model \cite{ymaosurvey} by defining the partition factor as $c_u(t) \in [0, 1]$, i.e., at the beginning of every time slot, the arrival task of each MID $u$ is divided into two disjoint parts. One part $A_u^l(t)=c_u(t)A_u(t)$ is executed locally while the other part $A_u^o(t)=(1-c_u(t))A_u(t)$ is offloaded to the associated MEC server in time slot $t$. By offloading all or partial computation tasks to a MEC server, the energy consumption of MID can be saved and the latency can be reduced as well. Since the partial offloading model is adopted, the offloading decisions in the multi-server multi-user MEC-enabled IoT systems include two parts, namely which MEC server to offload and how much to offload.  

\subsection{Queuing Model}
The queuing model is illustrated in Fig. 1. Two task buffers are maintained at each MID, where $Q_u^l$ stores  tasks for local processing and $Q_u^o$ stores tasks for offloading. At $t$, $D_u^l(t)$ bits  are processed locally while $D_u^o(t)$ bits  are offloaded to the associated MEC server.
The arriving tasks at $t$ can be queued in the buffer for future processing. The corresponding  evolution equation of $Q_u^l$ and  $Q_u^o$ can represented as
\begin{alignat}{5}
		\setlength{\abovedisplayskip}{3pt}
	\setlength{\belowdisplayskip}{3pt}
&Q_u^l(t + 1) = {[Q_u^l(t) - D_u^l(t)]^ + } + {c_u}(t){A_u}(t),\\
&Q_u^o(t + 1) = {[Q_u^o(t) - D_u^o(t)]^ + } + (1 - {c_u}(t)){A_u}(t),
\end{alignat}
where ${\left[ x \right]^ + } = \max \{ x,0\}$.

\subsection{Local Computing}
The local CPU-cycle frequency of MID $u$ at $t$ with a maximum value $f_{\max}$ is denoted as $f_u(t)$. Thus, the local processing data at $t$ can be expressed as
\be
		\setlength{\abovedisplayskip}{3pt}
	\setlength{\belowdisplayskip}{3pt}
D_u^l(t)= \tau f_u(t)/L_u,
\ee
where $L_u$ is the computation intensity in CPU cycles per bit.

Accordingly, the energy consumption for local computing at MID $u$  is given by
\be
		\setlength{\abovedisplayskip}{3pt}
	\setlength{\belowdisplayskip}{3pt}
{E_u^l}(t) = \tau\kappa {f_u}{(t)^3},
\ee
where $\kappa$ is the effective switched capacitance depending on the chip architecture of device.

\subsection{Computation Offloading}
An FDMA scheme is adopted for offloading, i.e., multiple MIDs associated with the same BS/MEC transmit simultaneously via different frequency bandwidth. The channel power gain from MID $u$ to MEC $m$ is given as $H_{um}(t) = {h_{um}}(t){g_0}({d_0}/{d_{um}})^\theta$, where $g_0$ is the path-loss coefficient, $\theta$ is the path-loss exponent, ${h_{um}}(t)$ is a small-scale Rayleigh fading coefficient from MID $u$ to MEC server $m$, which is exponentially distributed with a unit mean, $d_0$ is the reference distance, and ${d_{um}}$ is the distance from MID $u$ to server $m$.

Let $\mathcal{U}_m(t)=\{1,2,...N_m(t)\}$ denote the set of MIDs associated with MEC server $m$ at time slot $t$, where  $N_m=\sum\limits_{u = 1}^U {x_{um}}(t) \le N_{max}, \forall  m\in \mathcal{M},t \in \mathcal{T}$. Thus, the bandwidth allocation vector for all the MIDs associated with MEC $m$ can be represented as $ \pmb{\alpha}_m$ $\triangleq [\alpha_{1m}(t),\alpha_{2m}(t),...,\alpha_{N_mm}(t)]$, which should be chosen from the feasible set $\mathcal{A}_m$, i.e., ${ \pmb{\alpha}_m \in \mathcal{A}}_m \triangleq \{ \pmb{\alpha}_m \in \mathbb{R}_{+}^\mathit{N_m}| \sum_{u=1}^{N_m} \alpha_{um}(t)\le 1 \}$ \cite{vector}.

According to the Shannon-Hartley formula, the transmit rate from MID $u$ to MEC $m$ is expressed as
\be
\setlength{\abovedisplayskip}{3pt}
\setlength{\belowdisplayskip}{3pt}
r_{um}(t)= \left\{ {\begin{array}{*{20}{c}}
		{\alpha_{um}(t)\omega{\log_2}(1+\frac{{{H_{um}}(t)p_u^{tx}(t)}}{\chi
				+ \alpha_{um}(t)\omega\sigma ^2}),} & {\alpha_{um}(t) > 0 },  \\
		{0,} & {\alpha_{um}(t) =0}.  
\end{array}} \right.\label{ofrate}
\ee
where $p_u^{tx}(t)$ is the transmit power of MID $u$ with the  maximum value $P_{\max }^{tx}$. $\omega$ is the total uplink bandwidth for offloading, ${\sigma ^{\rm{2}}}$ is the background noise variance, and the variable $\chi$ is the average inter-cell interference power to be assumed a constant by exploring intelligent interference management technique according to the cell sizes \cite{inter1}\cite{Wenchao}.

The available bits offloaded from MID $u$ to its associated MEC are expressed as
\be
	\setlength{\abovedisplayskip}{3pt}
\setlength{\belowdisplayskip}{3pt}
D_u^o(t)=\sum_{m=1}^M x_{um}r_{um}(t)\tau.
\ee
The corresponding offloading energy consumption of MID $u$  is
 \be
   \setlength{\abovedisplayskip}{3pt}
 \setlength{\belowdisplayskip}{3pt}
 E_u^o(t) = p_{u}^{tx}(t) \tau.
 \ee
Therefore, the total computed bits and the energy consumption of MID $u$ at $t$ are respectively given by 
\be
	\setlength{\abovedisplayskip}{3pt}
\setlength{\belowdisplayskip}{3pt}
D_u(t)=D_u^l(t)+D_u^o(t),
\ee
\be
E_u(t)=E_u^l(t)+E_u^o(t).
\ee

\section{Problem Formulation}
In this section, we first introduce the performance metric of EE for IoT networks. Then, an EE-delay tradeoff problem for the MEC-enabled IoT networks is formulated, followed by an analytical framework based on Lyapunov optimization.
\subsection{Performance Metrics}
We define the network EE  as the ratio of long-term average energy consumption of all the MIDs to the corresponding long-term average aggregate accomplished computation tasks \cite{mao-multiaccess},
\be
		\setlength{\abovedisplayskip}{3pt}
	\setlength{\belowdisplayskip}{3pt}
\eta _{EE}\triangleq \frac{{\mathop {\lim }\limits_{T \to \infty } \frac{1}{T}\sum\limits_{t = 0}^{T - 1} {\mathbb{E}\{ E(t)\} } }}{{\mathop {\lim }\limits_{T \to \infty } \frac{1}{T}\sum\limits_{t = 0}^{T - 1} {\mathbb{E}\{ D(t)\} } }} = \frac{{\bar E}}{{\bar D}},
\ee
where $E(t) = \sum\limits_{u=1}^U E_u(t)$ is the total energy consumption of all the MIDs at $t$, $D(t) = \sum\limits_{u=1}^U D_u(t)$ is the total computed  data of all the MIDs at $t$.  $\bar E$ and $\bar D$ denote the long-term time average  of $E(t)$ and $D(t)$, respectively. 

According to Little's Law \cite{little}, the average service delay is proportional to the average queue length of task buffer. Denote $Q_u(t)$ as the sum queue length of MID $u$ at $t$, i.e., $Q_u(t)=Q_u^l(t)+Q_u^o(t)$. Hence, the time-average sum queue length of all the MIDs  is given by
\be
		\setlength{\abovedisplayskip}{3pt}
	\setlength{\belowdisplayskip}{3pt}
\overline Q_{\Sigma}=\lim_{T \to +\infty } \frac{1}{T} \sum\limits_{t = 0}^{T - 1}\sum\limits_{u = 1}^{U} \mathbb{E}\{Q_u(t)\}.
\ee
Given a task arrival rate, the average service delay is calculated by $[(\overline Q_{\Sigma})/(\sum_{u=1}^U \lambda_u)]$.
\subsection{EE-Delay Tradeoff Problem}
In order to reveal the fundamental tradeoff between EE and delay, our aim is to minimize the network long-term average EE while maintaining queue stability. The parameter vector for optimization in time slot $t$ is defined as $\mathcal{O}(t)\buildrel \Delta \over=\{\mathbf{c}(t), \pmb{\alpha}(t),\mathbf{x}(t),\mathbf{p}^{tx}(t),\mathbf{f}(t)\}$. The EE-delay tradeoff problem is formulated as 
\begin{subequations}
			\setlength{\abovedisplayskip}{3pt}
		\setlength{\belowdisplayskip}{3pt}
	\label{P1}
	\begin{alignat}{5}
	\textbf{P}{_1:}~~~ &\mathop {\min}_{\mathcal{O}(t)} \eta_{EE}\nonumber,\\	
	s.t.~~	& x_{um}\in \{ 0,1\} ,\forall m \in \mathcal{M},u \in \mathcal{U}, t \in \mathcal{T},\\
	&\sum\limits_{m = 1}^M { {x_{um}} \le 1,\forall u \in \mathcal{U}},t \in \mathcal{T},\\
	&N_m=\sum\limits_{u = 1}^U {x_{um}}(t) \le N_{\max}, \forall  m\in \mathcal{M},t \in \mathcal{T},\\
	& \alpha_{um}(t) \in \ \pmb{\alpha}_m \in \mathcal{A}_m,\forall m \in \mathcal{M},u \in \mathcal{U}_m(t), t \in \mathcal{T},\\
	&0 \le {c_u}(t) \le 1,\forall u \in \mathcal{U}, t \in \mathcal{T}, \\
    &0 \le {f_u}(t) \le {f^{\max }}, \forall u \in \mathcal{U}, t \in \mathcal{T}\\
    &0 \le {p_u^{tx}}(t) \le {P_{\max }^{tx}},\forall u \in \mathcal{U}, t \in \mathcal{T},\\   
	& \mathop {\lim }\limits_{t \to  + \infty } \frac{{\mathbb{E}\{ Q_u^l(t)\} }}{t} = 0,\forall u \in \mathcal{U}, t \in \mathcal{T},\\
	& \mathop {\lim }\limits_{t \to  + \infty } \frac{{\mathbb{E}\{ Q_u^o(t)\} }}{t} = 0, \forall u \in \mathcal{U}, t \in \mathcal{T}.
	\end{alignat}
\end{subequations}
The constraints in the formulation above can be explained as follows.
(\ref{P1}a) and (\ref{P1}b) imply that each MID can only offload its partial task to at most one MEC server on one frequency channel. (\ref{P1}c) indicates that the number of MIDs served by one MEC server concurrently in each time slot must not exceed the maximum allowable number. (\ref{P1}d) is the bandwidth allocation constraint,  (\ref{P1}e)-(\ref{P1}g) are the task partition factor constraint, the local CPU-cycle frequency constraint, and the allowable transmit power constraint, respectively.  (\ref{P1}h) and (\ref{P1}i) guarantee the task buffers $Q_u^l$ and $Q_u^o$ to be mean rate stable \cite{stochastic}.

Note that the original problem $\textbf{P}_1$ is an NP-hard problem due to the mixed-integer issues, the fractional structure, and the coupling of different optimization variables. Thus, it cannot be directly solved in a polynomial time. Therefore, we need to first transform it into a linear form by introducing an approximate variable based on the nonlinear fractional programming \cite{fracpro} to make the problem feasible. The similar method has been proved reasonable and widely adopted in \cite{mao-multiaccess, mobility-cache, ee2,ee3}.

 
\textit{Lemma 1:} The optimal network EE of $\text{P}{_1}$ can be equivalently rewritten as $ \min\bar{E}-\eta_{EE}^*\bar{D}$.

\textit{Proof:} The proof is similar with the proof in \cite{mobility-cache}, which is omitted here for brevity.

Since $\eta_{EE}^*$ is unknown in advance, to tackle the problem, we introduce a variable ${\eta _{EE}}(t)$, which is defined as
\be
\eta _{EE}(t) \triangleq{\frac{\sum\limits_{\upsilon=0}^{t-1} {E(\upsilon)}}{\sum\limits_{\upsilon=0 }^{t-1} {D(\upsilon)} } } \label{eta}.
\ee
Here, $\eta _{EE}(0)=0$,  and $\eta_{EE}(t)$ is a parameter that depends on the resource allocation strategies before $t$-th time slot. Through replacing  $\eta_{EE}^*$ as ${\eta _{EE}}(t)$ in $ \min\bar{E}-\eta_{EE}^*\bar{D}$,  we reformulate the original minimization problem  $\text{P}_1$ as  
\begin{subequations}
	\label{P2}
	\begin{alignat}{5}
	\textbf{P}{_2:}~~~ &\mathop {\min}_{\mathcal{O}(t)} \bar{E}-\eta_{EE}(t)\bar{D}, \nonumber\\	
	s.t.~~	&  (\ref{P1}a)- (\ref{P1}i)  \nonumber.
	\end{alignat}
\end{subequations}


\subsection{Lyapunov Optimization Framework}
To solve the stochastic optimization problem $\textbf{P}_2$, the Lyapunov optimization theory is explored in this section. 
First, we concentrate on the current queue backlogs  $\mathbf{\Theta}(t)=(\mathbf{Q}^o(t),(\mathbf{Q}^l(t))$, and introduce a quadratic Lyapunov function as 
\be
L(\mathbf{\Theta} (t))\mathop  = \limits^\Delta  \frac{1}{2}\sum_{u=1}^U \left[Q_u^l{(t)^2+Q_u^o(t)^2} \right].
\ee

The one-step conditional Lyapunov drift function is further introduced to push the quadratic Lyapunov function towards a bounded level and then stabilize the queue level.
\be
\Delta (\mathbf{\Theta} (t))\mathop  = \limits^\Delta  \mathbb{E}[L(\mathbf{\Theta} (t + 1)) - L(\mathbf{\Theta} (t))|\mathbf{\Theta} (t)].
\ee

By incorporating queue stability into the EE-delay tradeoff, we define a Lyapunov drift-plus-penalty function \cite{stochastic} as 
\be
{\Delta _V}(\mathbf{\Theta} (t)) = \Delta (\mathbf{\Theta} (t)) + V \cdot \mathbb{E}\left[E(t)-\eta_{EE} D(t)|\mathbf{\Theta} (t)\right], \label{drift}
\ee
where $V$ is the penalty weight of the objective function to the drift and can be also regarded as a control parameter to tune the EE and queuing delay.  In each time slot, by minimizing  ${\Delta _V}(\mathbf{\Theta} (t))$ (20), the optimal system EE can be achieved while the data queue level of MID $u$ can be stabilized. However, instead of directly minimizing  ${\Delta _V}(\mathbf{\Theta} (t))$ (20),  which is non-linear and intractable, we design the online offloading algorithm to minimize the upper bound of  ${\Delta _V}(\mathbf{\Theta} (t))$ according to \cite{stochastic}. Lemma 2 gives the upper bound of ${\Delta _V}(\mathbf{\Theta}(t))$ under any feasible $\mathcal O(t)$. 

\textit{Lemma 2:} In each time slot, for any queue backlogs $\mathbf{\Theta}(t)$,  the drift-plus-penalty function ${\Delta _V}(\mathbf{\Theta}(t))$ under an arbitrary feasible decision $\mathcal O(t)$ satisfies
\be
		\setlength{\abovedisplayskip}{3pt}
	\setlength{\belowdisplayskip}{3pt}
\begin{array}{l}
{\Delta _V}(\mathbf{\Theta}(t)) 
\le C_1 + V\mathbb{E}\left\{E(t)|\mathbf{\Theta}(t)\right\}\\
- \mathbb{E}\left\{\sum\limits_{u = 1}^U \left[V{\eta _{EE}}(t) + Q_u^l(t)\right]D_u^l(t)|\mathbf{\Theta}(t)\right\}\\
	 - \mathbb{E}\left\{\sum\limits_{u = 1}^U [V{\eta _{EE}}(t) + Q_u^o(t)]D_u^o(t)|\mathbf{\Theta}(t)\right\}\\
	  + \mathbb{E}\left\{\sum\limits_{u = 1}^U {Q_u^o(t){A_u}(t)}|\mathbf{\Theta}(t)\right\}\\   
+\mathbb{E}\Bigg\{\sum\limits_{u = 1}^U \left[ Q_u^l(t){c_u}(t) - Q_u^o(t){c_u}(t) + {c_u}(t)^2{A_u}(t) \right.   \\
\left. 	+ \frac{1}{2}{A_u}(t) - {c_u}(t){A_u}(t) \right]{A_u}(t)|\mathbf{\Theta}(t)\Bigg\},   \\ \label{driftpenalty}
\end{array}
\ee
	where $C_1=\frac{1}{2}\sum_{u=1}^U[(D^{l}_{u,\max})^2+(D^{o}_{u,\max})^2]
		\geq \frac{1}{2}\sum\limits_{u = 1}^U {[D_u^l{{(t)}^2} + D_u^o{{(t)}^2}} ]$, $D^{l}_{u,\max}$ and $D^{o}_{u,\max}$ are the upper bounds of $D_u^l(t)$ and $D_u^o(t)$, respectively. 

\textit{Proof:} Please refer to Appendix A.

Given the drift-plus-penalty bound found in Lemma 2, we design an online optimization approach based on Lyapunov optimization to minimize the right side of (\ref{driftpenalty}).
By doing so, the long term optimization problem $\text{P}_2$ is converted to a series of per-time slot based optimization problem $\text{P}{_3}$, which is much easier to handle
\be
\begin{aligned}
	\textbf{P}{_3:}~~  &\min_{\mathcal{O}(t)}~VE(t)- \sum\limits_{u = 1}^U \left[V{\eta _{EE}}(t) + Q_u^l(t)\right]D_u^l(t)\\
	&+\sum\limits_{u = 1}^U [ Q_u^l(t){c_u}(t) - Q_u^o(t){c_u}(t) + {c_u}(t)^2{A_u}(t)  \\
	&- {c_u}(t){A_u}(t) ]{A_u}(t) - \sum\limits_{u = 1}^U [V{\eta _{EE}}(t) + Q_u^o(t)]D_u^o(t) \\ 
   s.t.~~	&  (\ref{P1}a)- (\ref{P1}g)  \nonumber.
\end{aligned}
\ee

\section {Online Offloading and Resource Allocation Algorithm}
In this section, we study how to solve $\text{P}_3$, which mainly consists of three subproblems in each time slot: 1) task partition; 2) local computation resource allocation; 3) offloading resource allocation, which includes user association and radio resource allocation to determine the user association vector, transmit power and bandwidth factor.
We then introduce the proposed OORAA algorithm.

\subsection {Subproblem 1: Task Partition}

The optimal task partition factor at $t$ can be obtained by solving the following subproblem $\textbf{P}_{3.1}$:
\begin{subequations}
			\setlength{\abovedisplayskip}{3pt}
		\setlength{\belowdisplayskip}{3pt}
	\begin{alignat}{5}
	\textbf{P}_{3.1}: &\min_{\mathbf{c}(t)}\sum_{u=1}^U \left[ {c_u}{(t)^2}{A_u}(t) + {c_u}(t)(Q_u^l(t) - Q_u^o(t)- {A_u}(t)) \right]
	 \nonumber\\
	s.t.~~&	0 \le {c_u}(t) \le 1.
	\end{alignat}
\end{subequations}
$\textbf{P}_{3.1}$ is a convex problem with respect to $c_u(t)$ and the solution is given by

\be
	\setlength{\abovedisplayskip}{3pt}
\setlength{\belowdisplayskip}{3pt}
c_u^{* }(t) = \left\{ {\begin{array}{*{20}{c}}
		{0,} & {Q_u^o(t) \le Q_u^l(t) - {A_u}(t)},  \\
		{1,} & {Q_u^o(t) \ge Q_u^l(t) + {A_u}(t)},  \\
		{\frac{{Q_u^o(t) + {A_u}(t) - Q_u^l(t)}}{{2{A_u}(t)}},} &\text{otherwise.}
		\end{array}} \right.\label{fu33}
\ee

\subsection{Subproblem 2: Local computation resource allocation} 
By extracting the terms related to $\mathbf{f}(t)$, the optimal local CPU-cycle frequency scheduling problem is written as
\begin{subequations}
\begin{alignat}{5}
\textbf{P}_{3.2}:~ & \min_{\mathbf{f}(t)} -\sum\limits_{u = 1}^U {[Q_u^l(t) + V{\eta _{EE}}(t)]D_u^l(t) + } V\sum\limits_{u = 1}^U {{E_u^l}(t)} 
\nonumber\\	
s.t.~~&0 \le f_u(t) \le f^{\max}.
\end{alignat}
\end{subequations}

The optimal local CPU-cycle frequency can be calculated according to the stationary point and boundary value, which is given by
\be
	\setlength{\abovedisplayskip}{3pt}
\setlength{\belowdisplayskip}{3pt}
f_u^ * (t) = \min \left\{ \sqrt {\frac{{(Q_u^l(t) + V{\eta _{EE}}(t))}}{{3kV{L_u}}}} ,{f^{\max }}\right\}. 
\ee

\subsection{Subproblem 3: Offloading resource allocation}
The user association and radio resource bandwidth allocated to MIDs can be obtained by solving the following subproblem.

\begin{subequations}
			\setlength{\abovedisplayskip}{3pt}
		\setlength{\belowdisplayskip}{3pt}
\begin{alignat}{5}
	\textbf{P}_{3.3}{:} \min_{\mathbf{p}^{tx}(t), \pmb{\alpha}(t),\mathbf{x}(t)}
	&- \sum\limits_{u = 1}^U {(Q_u^o(t) + V{\eta _{EE}}(t))D_u^o(t) } \nonumber\\
	 &+V\sum\limits_{u = 1}^U {{p_u^{tx}(t)}\tau }\nonumber\\
	s.t.~~& (15a)-(15d),(15g).\nonumber
\end{alignat}
\end{subequations}

By replacing $\mathcal{A}_m$ with $\overline{\mathcal{A}}_m \triangleq \{ \pmb{\alpha}_m \in \mathbb{R}_{+}^\mathit{N_m}| \sum_{u=1}^{N_m} \alpha_{um}(t)\le 1, \alpha_{um}\ge \epsilon_m\}$, $\epsilon_m \in (0,1/N_{\max})$, a modified version  $\textbf{P}_{3.4}$ is formulated as
\begin{subequations}
	\setlength{\abovedisplayskip}{3pt}
\setlength{\belowdisplayskip}{3pt}
	\begin{alignat}{5}
	\textbf{P}_{3.4}{:} \min_{\mathbf{p}^{tx}(t), \pmb{\alpha}(t),\mathbf{x}(t)}
	&- \sum\limits_{u = 1}^U {(Q_u^o(t) + V{\eta _{EE}}(t))D_u^o(t) } \nonumber\\
	&+V\sum\limits_{u = 1}^U {p_u^{tx}(t)\tau }\nonumber\\
	s.t.~~& (15a)-(15d),(15g),\nonumber\\
	&\alpha_{m}(t)\in\overline{\mathcal{A}}_m.
	\end{alignat}
\end{subequations}

By doing so, we can transform $D_u^o(t)$ into a continuous and differentiable function with respect to $\alpha_{um}(t)$, which helps to develop the solution for $\textbf{P}_{3.4}$. By setting the value of $\epsilon_m$ sufficiently close to $0$, the optimal value of $\text{P}_{3.4}$ can be arbitrarily close to the original problem $\textbf{P}_{3.3}$ \cite{mao-tradeoff}.

It is evident that $\textbf{P}_{3.4}$ is a mixed-integer programming problem. The computational complexity is prohibitively high for a brute force approach.
It is noted that the feasible region is a Cartesian product of those of  $\pmb{\alpha}(t)$, $\mathbf{x}(t)$, $\mathbf{p}^{tx}(t)$. An alternating minimization procedure called Gauss-Seidel method \cite{seidel},\cite{yangyang} can effectively ensure the convergence. Motivated by that, we propose to solve $\textbf{P}_{3.4}$ alternately in an iterative way. 

\textbf{Optimal transmit power:} Assume  bandwidth allocation and association vector are given. 
Let ${\gamma _u}(t) = \sum\limits_{m = 1}^M {{\frac{x_{um}{H_{um}(t)}}{{\chi+\alpha_{um}(t)\omega{\sigma ^2} }}} }$, and ${B_u}(t) = (Q_u^o(t) + V{\eta _{EE}}(t))\alpha_{um}(t)\omega$. By decoupling all the MIDs, the problem $\textbf{P}_{3.4}$ can be converted into a set of the following problems, with each one corresponding to a different MID. 
\begin{subequations}
	\begin{alignat}{5}
			\setlength{\abovedisplayskip}{3pt}
		\setlength{\belowdisplayskip}{3pt}
	\textbf{P}_{3.4.1}{:} & \min_{{p}_u^{tx}(t)}
 - {B_u}(t)\tau {\log _2}(1 + {\gamma _u}(t)p_u^{tx}(t)) + Vp_u^{tx}(t)\tau \nonumber \\
	&s.t.~~ 0 \le p_u^{tx}(t) \le P_{\max }^{tx}.
	\end{alignat}
\end{subequations}

We denote the objective function of $\textbf{P}_{3.4.1}$ as $J(p_u^{tx}(t))$. The first and second derivative of $J(p_u^{tx}(t))$ with respect with $p_u^{tx}(t)$ are given by
\be
\begin{aligned}
		\setlength{\abovedisplayskip}{3pt}
	\setlength{\belowdisplayskip}{3pt}
	\frac{{d{J}({p_u^{tx}(t)})}}{{d{{p_u^{tx}(t)}}}}=\frac{{ - {B_u}(t)\tau {\gamma _u}(t) + V\tau (1 + {\gamma _u}(t)p_u^{tx}(t))\ln 2}}{{(1 + {\gamma _u}(t)p_u^{tx}(t))\ln 2}}
	\end{aligned}
\ee
\be
\begin{aligned}
		\setlength{\abovedisplayskip}{3pt}
	\setlength{\belowdisplayskip}{3pt}
		\frac{{d{J}^2({p_u^{tx}(t)})}}{{d{{p_u^{tx}(t)}}}^2}={B_u}(t)\tau \frac{{{\gamma _u}{{(t)}^2}}}{{{{(1 + {\gamma _u}(t)P)}^2}\ln 2}}.
\end{aligned}
\ee
It can be seen that $\frac{{d{J}^2({p_u^{tx}(t)})}}{{d{{p_u^{tx}(t)}}}^2} >0$, so  $\frac{{d{J}({p_u^{tx}(t)})}}{{d{{p_u^{tx}(t)}}}}$ always increases with $p_u^{tx}$. Suppose $\frac{{d{J}({p_u^{tx}(t)})}}{{d{{p_u^{tx}(t)}}}}=0$, one can get $p_u^{tx,0}(t) = \frac{{{B_u}(t)}}{{V\ln 2}} - \frac{1}{{{\gamma _u}(t)}}$. One can conclude that when $p_u^{tx}>p_u^{tx,0}$, $J(p_u^{tx}(t))$ is an increasing function of $p_u^{tx}(t)>0$; otherwise it is a decreasing function of  $p_u^{tx}(t)>0$. Therefore, the optimal transmit power of MID $u$ is calculated as
\be
		\setlength{\abovedisplayskip}{3pt}
	\setlength{\belowdisplayskip}{3pt}
P_u^{tx*}(t) = \left\{  \begin{array}{*{20}{c}}
0, & {\text{if}~V \ge \frac{{{B_u}(t){\gamma _u}(t)}}{{\ln 2}}}  \\
{\min \left\{ {P^{tx}_{\max }},\frac{{{B_u}(t)}}{{V\ln 2}} - \frac{1}{{{\gamma _u}(t)}}\right\}, } & \text{else}.  \\
\end{array}\right.
\ee

\textbf{Optimal Bandwidth Allocation:} 
For a given transmit power allocation and a given MEC association vector, the optimal bandwidth allocation can be obtained by solving the following subproblem.

\begin{subequations}
		\setlength{\abovedisplayskip}{3pt}
	\setlength{\belowdisplayskip}{3pt}
	\begin{alignat}{5}
	\textbf{P}_{3.4.2}{:}  \min_{\alpha_{um}}
	&-  (Q_u^o(t) + V{\eta _{EE}}(t))\sum_{m=1}^M\sum_{u=1}^{N_m} r_{um}(t)\tau \nonumber\\
	&s.t.~~ \epsilon_m \le\alpha_{um}(t), \forall u\in \mathcal{U}_m, \\
	&\sum_{u=1}^{N_m}\alpha_{um}(t)\le 1, \forall m\in \mathcal{M}.
	\end{alignat}
\end{subequations}

Since the bandwidth of a particular MEC is allocated to the MIDs that are associated with that MEC, and the bandwidth allocation of an MEC is independent of another MEC, $\textbf{P}_{3.4.2}$ can be further transformed into a series of bandwidth allocation problems by decoupling each MEC, given as
\begin{subequations}
		\setlength{\abovedisplayskip}{3pt}
	\setlength{\belowdisplayskip}{3pt}
	\begin{alignat}{5}
	\textbf{P}_{3.4.3}{:} & \min_{\alpha_{um}(t)}
	-  (Q_u^o(t) + V{\eta _{EE}}(t)) \sum_{u=1}^{N_m} r_{um}(t) \nonumber\\
	&s.t. (32a), (32b).\nonumber
	\end{alignat}
\end{subequations}

It's not difficult to find that  $\textbf{P}_{3.4.3}$ is convex, and the partial Lagrangian method can be exploited. The partial Lagrangian function of $\textbf{P}_{3.4.3}$  is obtained as 
\be
	\setlength{\abovedisplayskip}{3pt}
\setlength{\belowdisplayskip}{3pt}
	\begin{aligned}
	L(\alpha _{um}(t),\lambda_m(t))&=- (Q_u^o(t) + V{\eta _{EE}}(t))\sum\limits_{u = 1}^{N_m} {{r _{um}(t)}} \tau\\
	& + \lambda_m(t) (\sum\limits_{u = 1}^{N_m} {{\alpha _{um}}(t)}  - 1),  
	\end{aligned}
\ee
where $\lambda_m(t)\ge 0$ is the Lagrangian multiplier associated with $\sum_{u=1}^{N_m}\alpha_{um}(t)\le 1$.  The optimal bandwidth and the optimal Lagrangian multiplier
should satisfy the following equation set based on the Karush-Kuhn-Tucker (KKT) conditions.
\be
	\setlength{\abovedisplayskip}{3pt}
\setlength{\belowdisplayskip}{3pt}
\alpha _{um}^*(t) = \max \{\epsilon_m, g_{um}(\lambda_m^*(t)) \} ,
\ee 
where $g_{um}(\lambda_m^*(t))$ denotes the root of ${\frac{\partial L}{\partial \alpha _{um}(t)}}
=- [Q_u^o(t) + V{\eta _{EE}}(t)]\tau\frac{dr_{um}(t)}{d\alpha _{um}(t)}+\lambda_m(t)=0$ for $\lambda_m(t) >0$, $\frac{dr_{um}(t)}{d\alpha _{um}(t)}={{\omega} }[{\log _2}(1 + {\frac{{H_{um}}p_{um}^{tx}(t)}{\chi  + {\alpha _{um}}(t)\omega \sigma^2 }}) - {{{\alpha _{um}(t)}{H_{um}}p_{um}^{tx}(t)\omega \sigma^2 } \over  {\ln 2}{(\chi  + {\alpha _{um}}(t)\omega \sigma^2 )(\chi  + {\alpha _{um}}(t)\omega \sigma^2 + {H_{um}}p_{um}^{tx}(t))}}]$.

 As $\frac{dr_{um}(t)}{d\alpha _{um}(t)}$ is a decreasing function of $\alpha _{um}(t)$, $\mathop {\lim }\limits_{{\alpha _{um}}(t) \to \infty } {{d{r_{um}}(t)} \over {d{\alpha _{um}}(t)}} = 0$ and $\mathop {\lim }\limits_{{\alpha _{um}}(t) \to 0^+} {{d{r_{um}}(t)} \over {d{\alpha _{um}}(t)}} ={\omega } {\log _2}(1 + {{{H_{um}}p_{um}^{tx}(t)} \over {\chi}})\ge0 $, the root $g_{um}(\lambda_m)$ is positive and unique. Thus the bisection method over $[\lambda_m^L,\lambda_m^U]$ can be used for the search of the optimal $\lambda_m^*(t)$, where $\lambda_m^L=\min_{u\in\mathcal{U}_m}[Q_u^o(t) + V{\eta _{EE}}(t)]\tau\frac{dr_{um}(t)}{d\alpha _{um}(t)}|_{\alpha_{um}=1}$, $\lambda_m^U=\max_{u\in\mathcal{U}_m}[Q_u^o(t) + V{\eta _{EE}}(t)]\tau\frac{dr_{um}(t)}{d\alpha _{um}(t)}|_{\alpha_{um}=\epsilon_m}$ \cite{mao-tradeoff}. 
Furthermore, $g_{um}(\lambda_m^*(t))$ also can be obtained by a bisection search over $[\epsilon_m,1]$, and the searching process for the optimal $\alpha _{um}^*(t)$ is terminated when $|\sum_{u=1}^{N_m}\max\{\epsilon_m,g_{um}(\lambda_m(t))\}-1|\le\zeta$, where $\zeta$ is the error threshold. The details of the algorithm are presented as in Algorithm 1.
\begin{algorithm}[!t]
\setlength{\abovedisplayskip}{3pt}
\setlength{\belowdisplayskip}{3pt}
\algsetup{linenosize=\small}
\small
\caption{\textsc{Lagrangian Method for Bandwidth Allocation}}
\label{alg_graph1}
\begin{algorithmic}[1]
	\STATE \textbf{Initialization:}
	\STATE Set the error tolerant $\zeta=10^{-7}$, the temporary lower bound value $\lambda_m^{L,tmp}=\lambda_m^L(t)$, the temporary upper bound value $\lambda_m^{U,tmp}=\lambda_m^U(t)$, $\epsilon_m=10^{-4}$, the maximum iteration number $I=200$, $i=1$.
	\FOR {each MEC server}
   \FOR {$i\le I$}
	
	\STATE $\lambda_m^{tmp}=\frac{1}{2}(\lambda_m^{L,tmp}+\lambda_m^{U,tmp})$.
	\STATE	Set $\alpha_{um}^{(i)}(t)=\max \{\epsilon_m, g_{um}(\lambda_m^{tmp}) \}$
	\IF{$|\sum_{u=1}^{N_m}\alpha_{um}^{(i)}(t)-1|\le\zeta$}
	\STATE \textbf{break;}
	\ELSE
	\IF{$\sum_{u=1}^{N_m}\alpha_{um}^{(i)}(t)>1$}
	\STATE $\lambda_m^{L,tmp}=\lambda_m^{tmp}$
	\ELSE
	\STATE $\lambda_m^{U,tmp}=\lambda_m^{tmp}$
	\ENDIF
	\ENDIF
	\ENDFOR
	\ENDFOR
\end{algorithmic}
\end{algorithm}

\textbf{User Association:} 
For a given transmit power allocation and a given bandwidth allocation, the user association problem can be obtained by solving:
{\begin{subequations}
	\setlength{\abovedisplayskip}{3pt}
	\setlength{\belowdisplayskip}{3pt}
	\begin{alignat}{5}
	\begin{split}
	\textbf{P}_{3.4.4}{:} & \min_{\mathbf{x}(t)}
	- \sum_{u=1}^U\sum_{m=1}^Mx_{um}(t)\alpha_{um}(t)\omega\tau\\&~~~~{\log _2}\left(1
	+ \frac{H_{um}p_{um}^{tx}(t)}{\chi +{\alpha _{um}}(t)\omega\sigma^2}\right) 
	\end{split}\nonumber\\
	&s.t.~~ (15a)-(15c). \nonumber
	\end{alignat}
\end{subequations}}

Similar to \cite{yangyang,submodular}, we apply the submodular optimization here, which is regarded as a powerful tool for solving combinatorial problems.
Thus, in the following,  $\textbf{P}_{3.4.4}$ is first reformulated as a submodular function maximization problem under a matroid constraint. Then, a low-complexity greedy algorithm is utilized to obtain the optimal user association vector with the guaranteed performance.

The association action ground set is defined as as 
\be
\setlength{\abovedisplayskip}{3pt}
\setlength{\belowdisplayskip}{3pt}
\mathcal{G}(t) = \{ {\tilde{x}_{11}}(t), \ldots ,{\tilde{x}_{U1}}(t), \ldots ,{\tilde{x}_{1M}}(t),...,{\tilde{x}_{UM}}(t)\},  \nonumber
\ee
where $\tilde{x}_{um}(t)$ denotes the action that MID $u$ is associated with MEC server $m$ at $t$.

The ground set $\mathcal{G}(t)$ contains all possible association strategies which can be partitioned into $M$ disjoint sets, i.e., $\mathcal{G}(t) =  \cup _{m = 1}^M{A_m}(t), {A_m}(t) \cap {A_{m'}}(t) = \emptyset ,{\kern 1pt} {\kern 1pt} {\kern 1pt} {\kern 1pt} {\kern 1pt}$ for any $m \ne m'$. ${A_m}(t) = \{ \tilde{x}_{1m}^{}(t), \ldots ,\tilde{x}_{um}^{}(t)\},\forall m \in \mathcal{M}$, denoted the set of all MIDs that might associate with MID $m$ at $t$.

Given the finite ground set $\mathcal{G}(t)$, we define a partition matroid $(I_1; \mathcal{G}(t))$, where  $I_1 (t) \subseteq 2^{\mathcal{G}(t)}$ is a collection of independent sets given by
\be
\setlength{\abovedisplayskip}{3pt}
\setlength{\belowdisplayskip}{3pt}
{I_1}(t) = \{ \chi (t) \subseteq \mathcal{G}(t):|(\chi (t) \cap {A_m}(t))| \le {N_{\max }},\forall m \in M\},
\ee
which can be used to replace the constraint on the number of MIDs each MEC server serves at $t$ (c.f.(15c)). The set of MIDs associated with MEC server $m$ is denoted by $\chi_m(t)=\chi(t) \cap A_m(t)$.

The ground set can also be divided into $U$ disjoint sets, i.e., $\mathcal{G}(t) =  \cup _{u = 1}^U{B_u}(t)$, ${B_u}(t) \cap {B_{u'}}(t) = \emptyset  {\kern 1pt} {\kern 1pt} {\kern 1pt} {\kern 1pt} {\kern 1pt}$ for any $u \ne u'$, ${B_u}(t) = \{ \tilde{x}_{u1}^{}(t), \ldots ,\tilde{x}_{um}^{}(t)\}, \forall u \in \mathcal{U}$. Then we continue to define another partition matroid ($(I_2; \mathcal{G}(t))$), where the independent $I_2 (t) \subseteq 2^{\mathcal{G}(t)}$ is defined as
{\be
{I_2}(t) = \{ \chi (t) \subseteq \mathcal{G}(t):|\chi (t) \cap {B_u}(t)| \le 1, \forall u \in \mathcal{U}\},
\ee
which accounts for the constraint on the MEC server number each MID can associated with (c.f.(15.b)). The MEC server set associated with MID $u$ is denoted as $\chi_u(t)=\chi(t) \cap B_u(t)$.}

Let $W_{um}(t) =\alpha_{um}(t)\omega\tau {\log _2}(1 +\frac{{H_{um}(t)p_u^{tx}(t)}}{{\chi  + {\alpha _{um}(t)}\omega {\sigma ^2}}})$. By considering the partition matriod $I_1 (t) \subseteq 2^{\mathcal{G}(t)}$ and $I_2 (t) \subseteq 2^{\mathcal{G}(t)}$, the following theorem reveals the the property of the objective function in problem $\textbf{P}_{3.4.4}$.

{\textit{Theorem 1:} Given a ground set $\mathcal{G}(t)$ and a subset $\chi (t) \in \mathcal{G}(t)$, the set function $J(\chi (t)) = \sum\limits_{\tilde{x}_{um}(t) \in \chi (t)}^{} {W_{um}^{}(t)}$ is a monotone submodular function over $\chi (t) \in \mathcal{G}(t)$.}

\textit{Proof:} Please refer to Appendix B.

On the basis of theorem 1, problem  $\textbf{P}_{3.4.4}$ can be formulated as a monotone submodular maximization problem with two partition matroid constraints:
\begin{subequations}
	\setlength{\abovedisplayskip}{3pt}
	\setlength{\belowdisplayskip}{3pt}
	\begin{alignat}{5}
	\textbf{P}_{3.4.5}{:} ~~ &\max \limits_{\chi (t)} \sum\limits_{\tilde{x}_{um}(t) \in \chi (t)} W_{um}(t)\nonumber\\
	&s.t.~~\chi (t) \in {I_1}(t) \cap {I_2}(t).
	 \nonumber
	\end{alignat}
\end{subequations}

An effective approach towards maximizing a monotone submodular function in case of matroid constraints is the greedy algorithm\cite{Wenchao}, \cite{submodular}.
Thus we adopt a greedy algorithm to find suboptimal solution for problem $\textbf{P}_{3.4.5}$ shown in Algorithm 2.

Define the marginal gain of $J(\chi (t))$ by adding one element $\tilde{x}_{um}(t) \in  \mathcal{G}(t)\backslash\chi (t) $ to the set of $\chi (t)$ as
\be
	\setlength{\abovedisplayskip}{3pt}
\setlength{\belowdisplayskip}{3pt}
\triangle_{\chi(t)}(\tilde{x}_{um}(t))=J(\chi(t) \cup {\tilde{x}_{um}}(t)) - J(\chi(t))=W_{um}(t).
\ee

At first, $\chi(t)$, $\chi_m(t)$ and $\chi_u(t)$ are initialized as empty sets $\emptyset$, while $\mathcal{S}(t) $ is initialized as the set $\mathcal{G}(t)$. In each iteration, we calculate the {marginal gain} $\triangle_{\chi(t)}(\tilde{x}_{um}(t))$ for each element $\tilde{x}_{um}(t) \in \mathcal{G}(t) \backslash \chi (t)  $ and select the optimal element with the highest {marginal gain}, i.e., 
\be
	\setlength{\abovedisplayskip}{3pt}
\setlength{\belowdisplayskip}{3pt}
 \tilde{x}_{um}(t) = \mathop {\arg \max }\limits_{\tilde{x}_{um}(t) \in \mathcal{G}(t) \backslash \chi (t) } \triangle_{\chi(t)}(\tilde{x}_{um}(t)).
\ee
Then, we add this element $ \tilde{x}_{um}(t)$ into the set $\chi(t)$, the set $\chi_m(t)$ as well as the set $\chi_u(t)$. Since adding one element into $\chi_u(t)$ means that user $u$ has determined its single associated MEC server, then we remove the $B_u(t)$ from the set $\mathcal{S}$. When the set $\chi_m(t)$ has accumulated to $N_{max}$ elements, the set $A_m(t)$ is removed from the set $\mathcal{S}(t)$. The iterations continue  until no more element can be added., i.e., the marginal value $W_{um}(t)$  is zero for all $\tilde{x}_{um} \in \mathcal{S} \backslash \chi$ or $\mathcal{S}=\emptyset$.

\begin{algorithm}[!t]
	\setlength{\abovedisplayskip}{3pt}
	\algsetup{linenosize=\small}
	\small
	\caption{\textsc{Greedy Algorithm for Optimal Association }}
	\label{alg_graph}
	\begin{algorithmic}[1]
	\STATE \textbf{Initialization:}
	\STATE Set  $\mathcal{S}(t)=\mathcal{G}(t)$, $\chi (t) = \emptyset$, $\chi_u (t) =\emptyset$, $\chi_m (t) =\emptyset$;
		
		\STATE \textbf{Repeat} 
		\STATE	Find  the element $\tilde{x}_{um}(t)$ with the highest marginal gain,
		
		 $\tilde{x}_{um}(t) = \mathop {\arg \max }\limits_{\tilde{x}_{um}(t) \in \mathcal{S} \backslash \chi(t)} W_{um}(t)$
		\STATE \textbf{Set} $\chi (t) = \chi (t) \cup  \tilde{x}_{um}^{}(t) $, $\chi_m(t)=\chi_m(t) \cup \tilde{x}_{um}(t)$, $\chi_u(t)=\chi_u(t) \cup \tilde{x}_{um}(t)$;
		\STATE update $\mathcal{S}(t) = \mathcal{S}(t)\backslash {B_u}(t)$
	    \STATE \textbf{if} $|\chi_m(t)|=N_{max}$, \textbf{then}
	    \STATE  update $\mathcal{S}(t) = \mathcal{S}(t)\backslash {A_m}(t)$;
	    \STATE \textbf{end if}
		\STATE Go back to step 3
		 \STATE until $\mathcal{S}(t)=\emptyset$ or $W_{um}(t) = 0$ for all $\tilde{x}_{um}(t) \in \mathcal{S}(t) \backslash \chi(t)$

		\STATE \textbf{Output:} $\chi (t)$.
	\end{algorithmic}
\end{algorithm}

\subsection{{{OORAA}} Algorithm and Theoretic Analysis}
By jointly considering the dynamic user association, offloading decision, radio and computational resource allocation, the {OORAA} algorithm is summarized in Algorithm 3.

\begin{algorithm}[!t]
	\algsetup{linenosize=\small}
	\small
	\caption{\textsc{The Proposed OORAA Algorithm.}}
	\label{alg2}
	\begin{algorithmic}[1]
		\STATE 
	    At each beginning of time, for $\forall u\in U$, obtain $A_u(t)$,$H_{um}(t)$, $Q_u^l(t)$, $Q_u^o(t)$, $\eta_{EE}$, and the control parameter $V$.\\

		\WHILE {$t \le t_{end}$}
	    \STATE \textbf{Repeat:}
		\STATE Task-partition Decision:
		\STATE Determine $C_u^*(t)$ according to (23) for $\forall u\in U$.
		\STATE Local computational resource Scheduling: 
		\STATE Determine $f_u^*(t)$ according to (25) for $\forall u\in U$.
		\STATE Random initialize $x_{um}^{(0)}(t)$ and $\alpha_{um}^{(0)}(t)$.\\
		Set $\mathbf{x}^{(0)}(t)=\{x_{um}^{(0)}(t)\}$, $\pmb{\alpha}^{(0)}(t)=\{\alpha_{um}^{(0)}(t)\}$.
		\STATE \textbf{Initialization: $k=0$}
		\STATE \textbf{Repeat:}
		\STATE With given $\mathbf{x}^{(k)}(t)$,  $\pmb{\alpha}^{(k)}(t)$, compute $p_u^{tx,(k+1)}(t)$ by (31) . Set $\mathbf{p}^{tx,(k+1)}=\{p_u^{tx,(k+1)}(t)\}$
		\STATE 	With given $\mathbf{x}^{(k)}(t)$, $\mathbf{p}^{tx,(k+1)}$, compute $\alpha_{um}^{(k+1)}(t)$, $\forall u\in U_m$ by Algorithm 1 .
		\STATE With given $\mathbf{p}^{tx,(k+1)}$, $\pmb{\alpha}^{(k+1)}(t)$, compute $x_{um}^{(k+1)}(t)$ by Algorithm 2.
		\STATE 	Output $\mathbf{p}^{tx,(k+1)}$,$\pmb{\alpha}^{(k+1)}(t)$, $\mathbf{x}^{(k+1)}(t)$for next iteration.
		\STATE Set $k=k+1$.
		\STATE Until convergence
		\STATE $\mathbf{x}^{*}(t)$, $\pmb{\alpha}^{*}(t)$, and $\mathbf{p}^{tx*}(t)$
		
		\STATE Queue Length Update:
		\STATE Update the local task queue according to (3);
		\STATE Update the offloading task queue according to (4);
		\STATE Update the EE according to (12). 
		\STATE Set $t=t+1$.
		\ENDWHILE
	\end{algorithmic}
\end{algorithm}

In the following, we present the performance analysis of the proposed OORAA algorithm, including the queue stability, the upper bounds of EE and sum queue lengths.

Lemma 3 shows that there exist stationary and randomized policies for ${\text{P}_1}$ \cite{stochastic},
based on which, the offloading decisions and resource scheduling can be determined independently among different time slots and depend only on $A_i(t)$. 

\textit{Lemma 3: } Suppose the original problem ${\textbf{P}_1}$ is strictly feasible for a set $\{\lambda_u\}$. Thus, there exists a positive value $\epsilon$ to make the problem  ${\textbf{P}_1}$ feasible for $\{\lambda_u+\epsilon\}$. Therefore, for any $\delta >0$, there exists an independent, stationary and randomized policy, which satisfies the following inequalities:
\begin{alignat}{5}
		\setlength{\abovedisplayskip}{3pt}
	\setlength{\belowdisplayskip}{3pt}
& \label{eq1} \mathbb{E}\{E^{\Pi}(t)\} \le \mathbb{E} \sum\limits_{u = 1}^U \left[D_u^{l,\Pi}(t)+D_u^{o,\Pi}(t)\right]
(\eta_{EE}^*+\delta),  \\
& \label{eq2} \mathbb{E}\{D_u^{l,\Pi}(t)-c_u^\Pi(t)A_i(t)|\mathbf{\Theta}(t)\}=\mathbb{E}\{D_u^{l,\Pi}(t)\}-c_u^\Pi(t)\lambda_u \ge \epsilon,\\   
& \label{eq3} \mathbb{E}\{D_u^{o,\Pi}(t)-(1-c_u^\Pi(t))A_i(t)|\mathbf{\Theta}(t)\}\\  \nonumber 
&~~~~~=\mathbb{E}\{D_u^{o,\Pi}(t)\}-(1-c_u^\Pi(t))\lambda_u \ge \epsilon,  
\end{alignat}
where $E^{\Pi}(t)$, $D_u^{l,\Pi}(t)$, $D_u^{o,\Pi}(t)$, and $c^\Pi _u(t)$ are the resulting values under the independent, stationary, and randomized algorithm; $\eta_{EE}^*$ is the optimal solution of problem $\textbf{P}_1$.

\textit{Proof:} The detailed proof is omitted for brevity, a similar proof can be found in \cite{stochastic}.

By leveraging Lemma 3, the following theorem presents its asymptotic optimality for the original problem ${\textbf{P}_1}$. 

\textit{Theorem 2:} Suppose the original EE-delay tradeoff problem $\textbf{P}_1$ is strictly feasible, $\mathbb{E}\{L(\Theta(0))\} \le \infty$. Then for any $t \in \mathcal{T}$,  $V>0$, $\epsilon>0$, the relationship between network EE $\eta_{EE}$ obtained by the proposed OORAA algorithm and the optimal value $\eta_{EE}^*$ of $\textbf{P}_1$ is given as
\be
	\setlength{\abovedisplayskip}{3pt}
\setlength{\belowdisplayskip}{3pt}
\eta_{EE}\le\eta_{EE}^*+\frac{C_1+C_2}{V(D^{l}_{\min}+D^{o}_{\min })}.
\ee 
The time-average sum data queue length of all the MIDs is bounded by
\be
	\setlength{\abovedisplayskip}{3pt}
\setlength{\belowdisplayskip}{3pt}
\overline Q_{\Sigma} \le \frac{{C_1+C_2 + V\eta _{EE}^ * (D^{l}_{\max } + D^{o}_{\max }) - {E^{\min }}}}{\epsilon },
\ee
where $ C_2=\frac{A^2_{max}}{2}$,  $ D^{l}_{\min}=\min(\{\sum\limits_{u=1}^{U}D_u^l(t)\})$, $D^{o}_{\min}=\min(\{\sum\limits_{u=1}^{U}D_u^o(t)\})$, $ D^{l}_{\max}=\max(\{\sum\limits_{u=1}^{U}D_u^l(t)\})$, $ D^{o}_{\max}=\max(\{\sum\limits_{u=1}^{U}D_u^o(t)\})$, and $E^{min}$ is the minimum energy consumption.

\textit{Proof:} Please refer to Appendix C.

\textit{Remark 1}: Theorem 2 demonstrates that the proposed OORAA algorithm achieves a $[O(1/V), O(V)]$ tradeoff between EE and queue backlog. As mentioned before, the network service delay is proportional to the time-averaged data queue length, the same tradeoff also exists between EE and the service delay.
In addition, the optimal EE of the original problem can be approached by letting $V \rightarrow\infty $.

In each per-time slot problem  $\text{P}_{3}$, the computational complexity of solving  $\text{P}_{3.1}$ and $\text{P}_{3.2}$ is $2U$. An iterative way is utilized to solve $\text{P}_{3.4}$ with $K$ maximum iterations. In each iteration, the computational complexity of solving  $\text{P}_{3.4.1}$ is $U$. Then the Lagrangian method is employed to solve $\text{P}_{3.4.2}$, which employs bisection search methods to find the optimal value of $\lambda_m^*(t)$ and 
$\alpha _{um}^*(t)$ with $\log_2\left(\frac{\lambda_U(t)-\lambda_L(t)}{\lambda_{\zeta}(t)}\right)$ and $\log_2(\frac{1}{\epsilon})$, respectively,  where $\lambda_{\zeta}(t)$  corresponds to the accuracy requirement for $\lambda_m^*(t)$ given $\zeta$, $\epsilon$ corresponds to the accuracy requirement for $\alpha _{um}^*(t)$. Thus, the total complexity for algorithm 1 is $O(MN_{max}(\log_2\left(\frac{\lambda_U(t)-\lambda_L(t)}{\lambda_{\zeta}(t)}\right)\log_2(\frac{1}{\epsilon})))$ \cite{mao-tradeoff}. Moreover, since there are $UM$ elements in the ground set $\mathcal{G}(t)$, the time cost to solve $\text{P}_{3.4.2}$ in algorithm 2 is given as $O(UM)$ \cite{submodular1}. Therefore, the total complexity of the proposed algorithm OORAA (algorithm 3) for each time slot is given as $2U+K[U+MN_{max}(\log_2\left(\frac{\lambda_U(t)-\lambda_L(t)}{\lambda_{\zeta}(t)}\right)\log_2(\frac{1}{\epsilon}))+UM]$.
\section{Simulation Results}
\label{Simulation}
In this section, simulation results are provided to validate the theoretical analysis and evaluate the performance of the proposed {OORAA algorithm}. The simulation settings are based on the work in \cite{mao-energy} and \cite{mao-tradeoff}. We consider a scenario that multiple MECs and multiple MIDs are randomly deployed in a $100\times 100~m^2$ area. Each MEC server serves at most $N_{\max}=4$ MIDs at one time. The MID trajectory is generated by using the random walk model in \cite{mobilitysun}.  
The channel power gains are exponentially distributed with the mean of $g_0\cdot (d/d_0)^{-4}$, where the reference distance $d_0=1$ and $g_0 =-40$ dB. The arrival task size $\lambda_u(t)$ in every time slot is uniformly distributed within $[1, 2]\times 10^3$ bits. 
In addition, $M=3$, $U=10$, $\kappa = 10^{-28}$, $\omega= 1$ MHz,  $\sigma^2 = -174$ dBm/Hz, $P_{\max}^{tx} = 1$ W, $f^{\max} = 2.15$ GHz, $\gamma_u=737.5$ cycles/bit,  $\chi=10^{-13}W$  and the time slot duration $\tau=1$ ms unless otherwise specified.

\subsection{Performance and Parameter Analysis}

\begin{figure}[h]
	\vspace{-0.3cm}
	\setlength{\abovecaptionskip}{-0.2cm} 
	\setlength{\belowcaptionskip}{-1cm}
	\centering
	\setlength{\abovecaptionskip}{-0.2cm} 
	\setlength{\belowcaptionskip}{-1cm}
	\centering
	\includegraphics[width=3.0in]{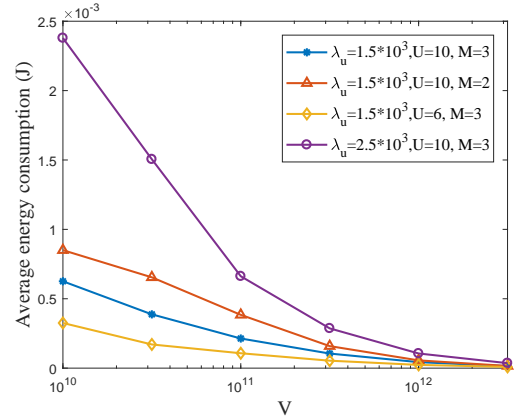}	
	\caption{Average energy consumption v.s. control parameter $V$.\label{v-ene}}
\end{figure} 

\begin{figure}[h]
	\vspace{-0.3cm}
	\setlength{\abovecaptionskip}{-0.2cm} 
	\setlength{\belowcaptionskip}{-1cm}
	\centering
		 \vspace{-0.3cm}
		 \setlength{\abovecaptionskip}{-0.2cm} 
		 \setlength{\belowcaptionskip}{-1cm}
		\centering
	\includegraphics[width=3.0in]{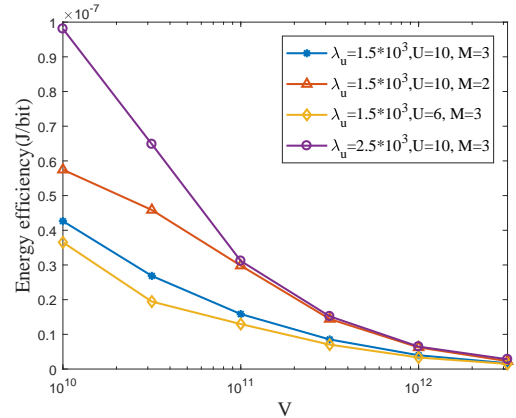}	
\caption{EE v.s. control parameter $V$.\label{v-ee}}
\end{figure}

\begin{figure}[h]
	\vspace{-0.3cm}
	\setlength{\abovecaptionskip}{-0.2cm} 
	\setlength{\belowcaptionskip}{-1cm}
	\centering
	\setlength{\abovecaptionskip}{-0.2cm} 
	\setlength{\belowcaptionskip}{-1cm}
	\centering
	\includegraphics[width=3.0in]{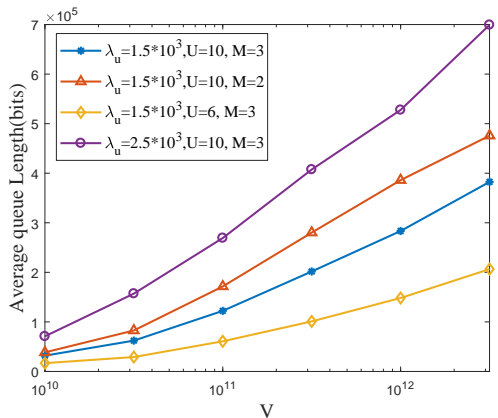}	
	\caption{Average Queue length v.s. control parameter $V$.\label{v-Q}}
\end{figure}

\begin{figure}[h]
	\vspace{-0.3cm}
	\setlength{\abovecaptionskip}{-0.2cm} 
	\setlength{\belowcaptionskip}{-1cm}
	\centering
	\setlength{\abovecaptionskip}{-0.2cm} 
	\setlength{\belowcaptionskip}{-1cm}
	\centering
	\includegraphics[width=3.0in]{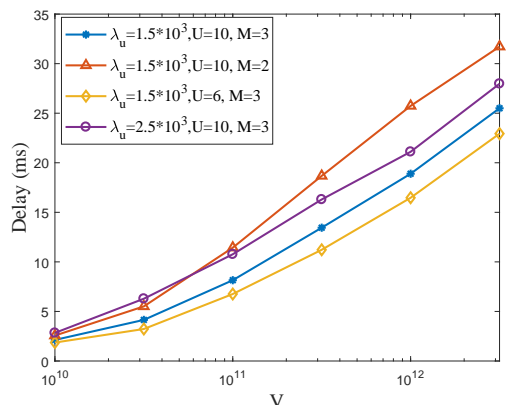}	
	\caption{Delay v.s. control parameter $V$.\label{v-delay}}
\end{figure}

The performance of energy consumption vs. the control parameter is presented in Fig. \ref{v-ene}. The energy consumption of all the settings  decreases with the increase of control parameter $V$. This is due to the fact that a larger value of $V$ makes the system more inclined  to  save the energy consumption, which agrees with the definition of the objective function $\textbf{P}_3$. In addition, the difference among all the curves can be explained as follows. Since increasing the task arrival rate or the number of MIDs requires a higher task processing rate to maintain the target queue size, the energy consumption increases accordingly. Moreover, when the number of MEC servers decreases from $3$ to $2$, the average number of MIDs associated with one MEC increases, which gives less bandwidth to each associated MID.  

Fig. \ref{v-ee} shows the relationship between EE and the control parameter $V$. We can observe that the network EE decreases as $V$ increases, and converges to the optimal value $\eta_{EE}^*$ when $V$ is sufficiently large, which verifies the asymptotic optimality developed in Theorem 1. Given $M$ MEC servers, EE increase with the  growth of the task arrival rate and the increase of the number of MIDs.  A larger task arrival rate and a greater number of MIDs cause the increase of the average energy consumption in order to keep the queue stable. Meanwhile, according to (6)-(7) and (9)-(10), one  can find that when the data rate is high enough, the energy consumption goes up faster  than the task completion rate, leading to a decrease in EE. 

Fig. \ref{v-Q} and Fig. \ref{v-delay} show the impact of the control parameter $V$ on the average queue length and the service delay, respectively. Since the average service delay is proportional to the average queue length, the trend of average length and the trend of the average delay are the same.
It can be observed that the average length and average delay both increase with the control parameter $V$, which is in conformity with Theorem 1. In addition, a higher task arrival rate and a larger MID number cause the increase of the average queue length and the average service delay, which is easy to understand. Moreover, given the task arrival rate and the MID number, the average queue length and the average delay decrease with the increase of the number of MEC servers. As the number of MEC servers increases, the offloading bandwidth provided for each MID may increase. Thus the tasks buffered at the offloading queue can be processed rapidly, which then reduces the average queue length and the corresponding queuing delay of the MID.

\begin{figure}[h]
	\vspace{-0.3cm}
	\setlength{\abovecaptionskip}{-0.2cm} 
	\setlength{\belowcaptionskip}{-1cm}
	\centering
	\setlength{\abovecaptionskip}{-0.2cm} 
	\setlength{\belowcaptionskip}{-1cm}
	\centering
	\includegraphics[width=3.0in]{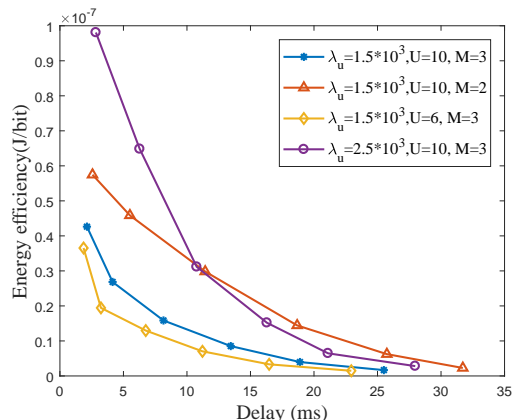}	
	\caption{EE v.s. average delay .\label{delay-ee}}
\end{figure}

Fig. \ref{delay-ee} depicts the network EE with respect to the average service delay. 
It is quite evident that the network EE decreases when the average service delay increases. This is due to the fact that there exist a intrinsic tradeoff between EE and delay on devices, i.e., a small CPU-cycle frequency or transmit power can save energy consumption but incur the long completion latency. We can also see that the network EE is inversely related to the average service delay, further demonstrate the $[O(1/V), O(V)]$ tradeoff between the EE and the service delay, in which the network EE decrease with $V$ while the average delay increases with $V$. Therefore, $V$ is a critical parameter to balance the EE and the average delay, which should be chosen carefully according to different aims in practical system design.

\begin{figure}[h]
	\vspace{-0.3cm}
	\centering
	\includegraphics[width=3.0in]{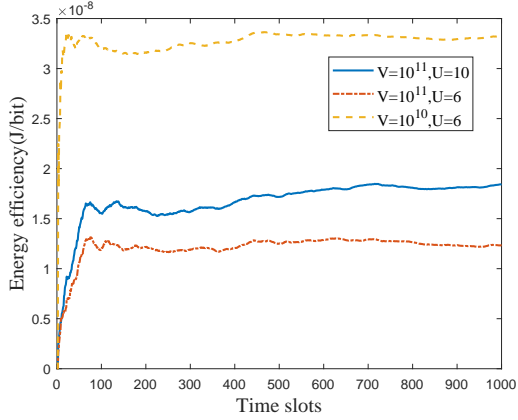}	
	\caption{EE v.s. time slot.\label{syslot}}
\end{figure}

The network EE with respect to time is shown in Fig. \ref{syslot}, from which we can observe that the proposed algorithm converges to a stable performance when time evolves. It further demonstrates that a larger value of $V$ or a smaller number of MIDs can result in lower energy efficiency, which aligns with the observations in Fig. \ref{v-ee}.

\subsection{Comparison With Other Schemes}

In this subsection,  the proposed algorithm is compared with the other four benchmark schemes, namely, `Complete local processing', `Complete offloading', `Random 0-1 offloading', and `Random association'. The control parameter $V$ in the following simulations is set to $10^{11}$. The benchmark schemes are defined as follows:

1) Complete local processing: All the  computation tasks are executed locally, i.e., $c_u(t)=0, \forall u \in \mathcal{U} $, $t \in \mathcal{T}$.

2) Complete offloading: All the computation tasks are offloaded to the remote MEC server and there is no local processing, i.e., $c_u(t)=1, \forall u \in \mathcal{U} $, $t \in \mathcal{T}$. 

3) Random 0-1 offloading: The computation tasks are either executed locally or offloaded to the associated MEC server. The decision is made by each MID randomly. 

4) Random association: Each MID randomly selects the associated MEC but satisfies the constraint that the number of MIDs associated with one MEC server cannot exceed $N_{\max}$. 

\begin{figure}[h]
	\centering
	\begin{minipage}[t]{0.48\textwidth}
		\centering
		\includegraphics[width=3.0in]{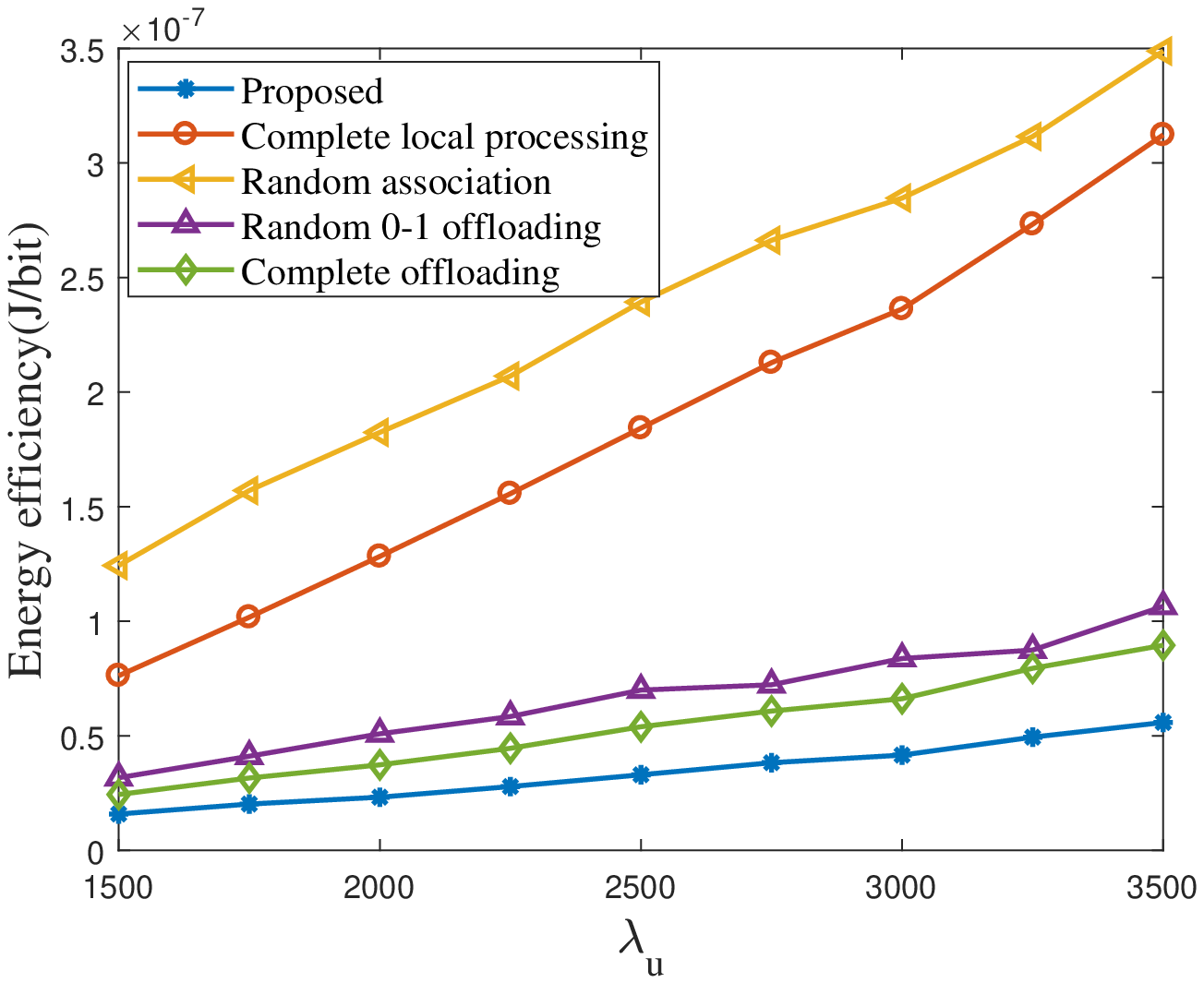}	
		\caption{EE v.s. average task size $\lambda_u$.\label{lambda-ee}}
	\end{minipage}
	\begin{minipage}[t]{0.48\textwidth}
		\centering
		\includegraphics[width=3.0in]{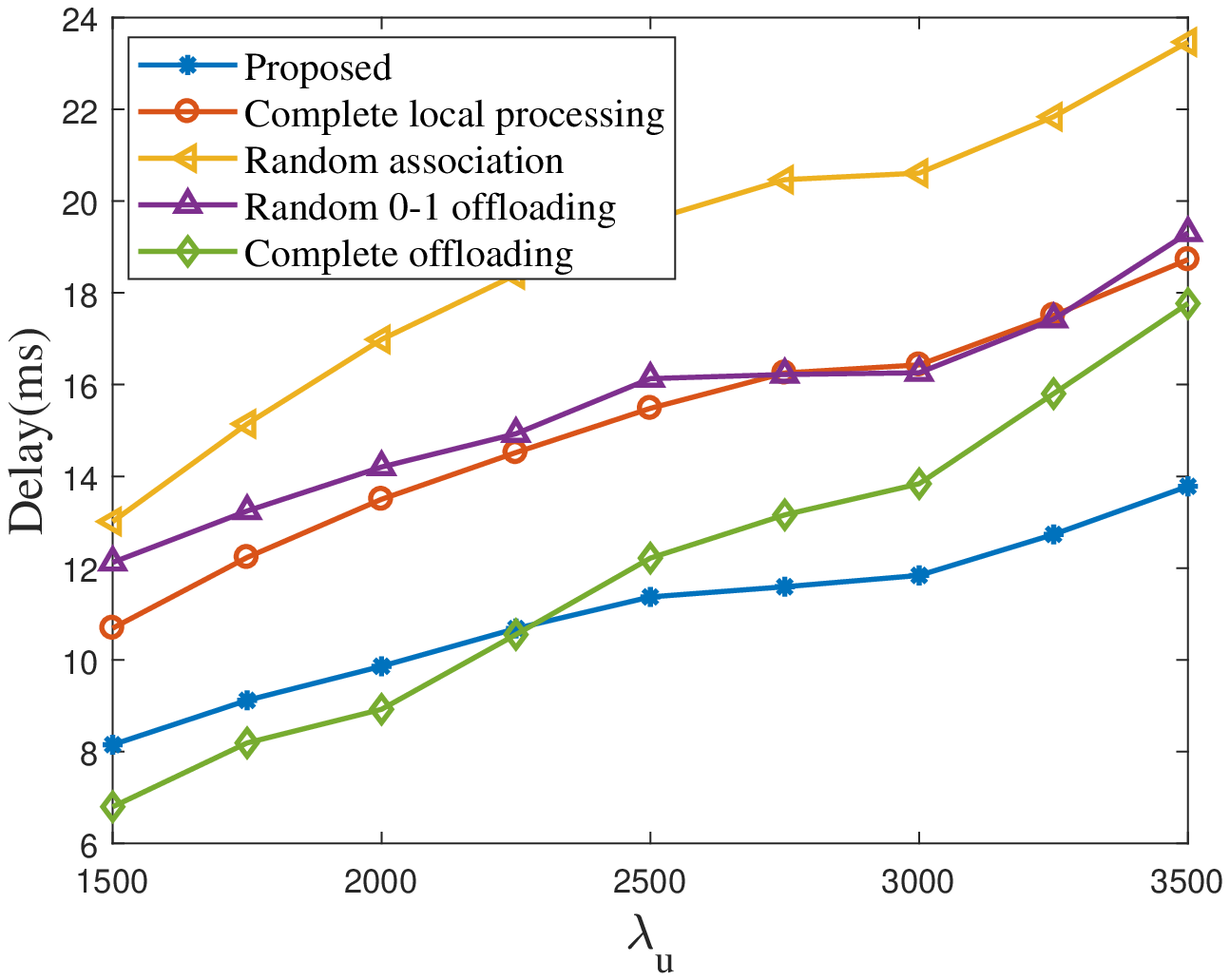}	
		\caption{ Delay v.s. average task size $\lambda_u$.\label{lambda-delay}}
	\end{minipage}
\end{figure}

\begin{figure}[h]
	\centering
	\begin{minipage}[t]{0.48\textwidth}
		\centering
		\includegraphics[width=3.0in]{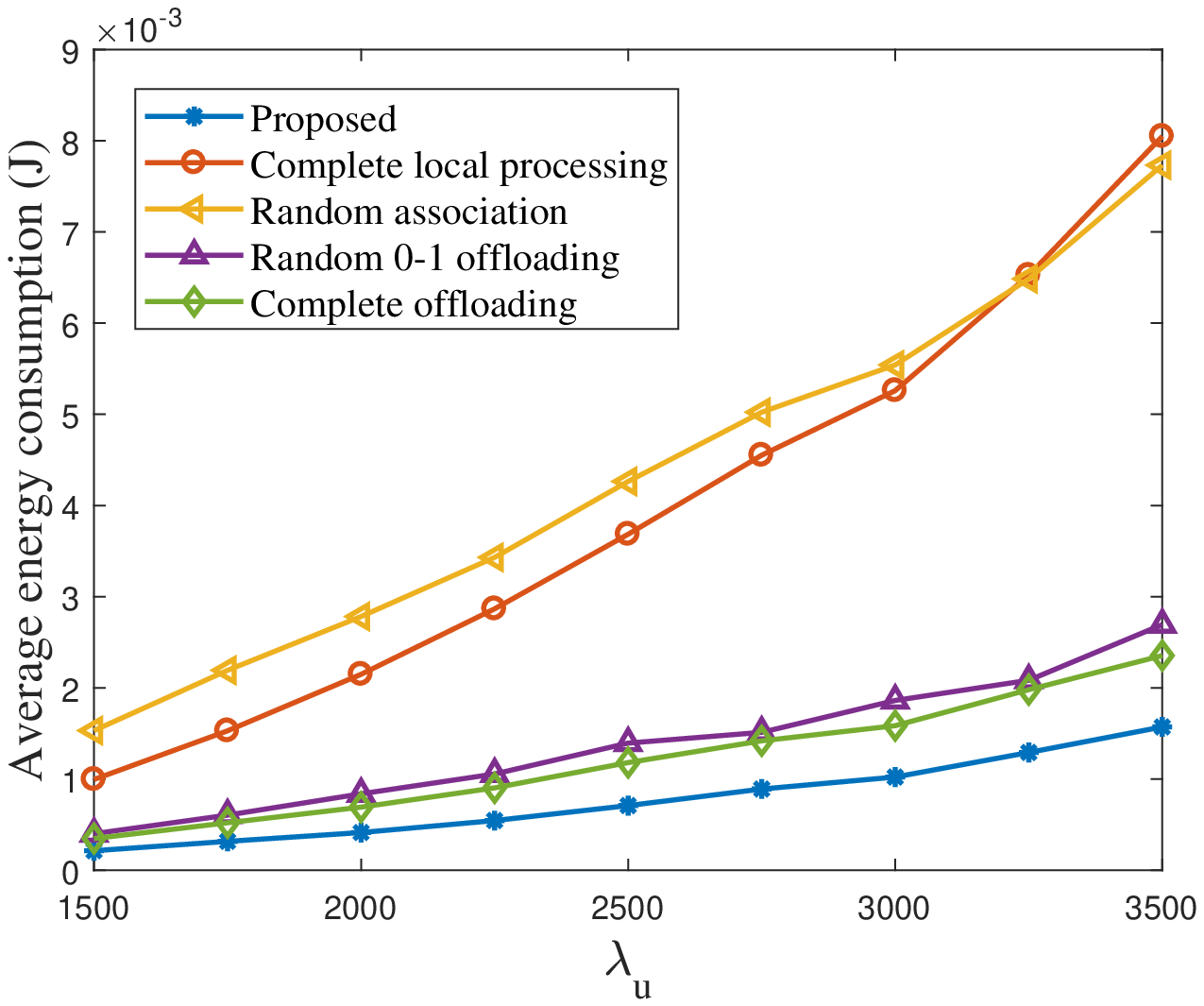}	
		\caption{Energy consumption v.s. average task size $\lambda_u$.\label{lambda-ener}}
	\end{minipage}
	\begin{minipage}[t]{0.48\textwidth}
		\centering
		\includegraphics[width=3.0in]{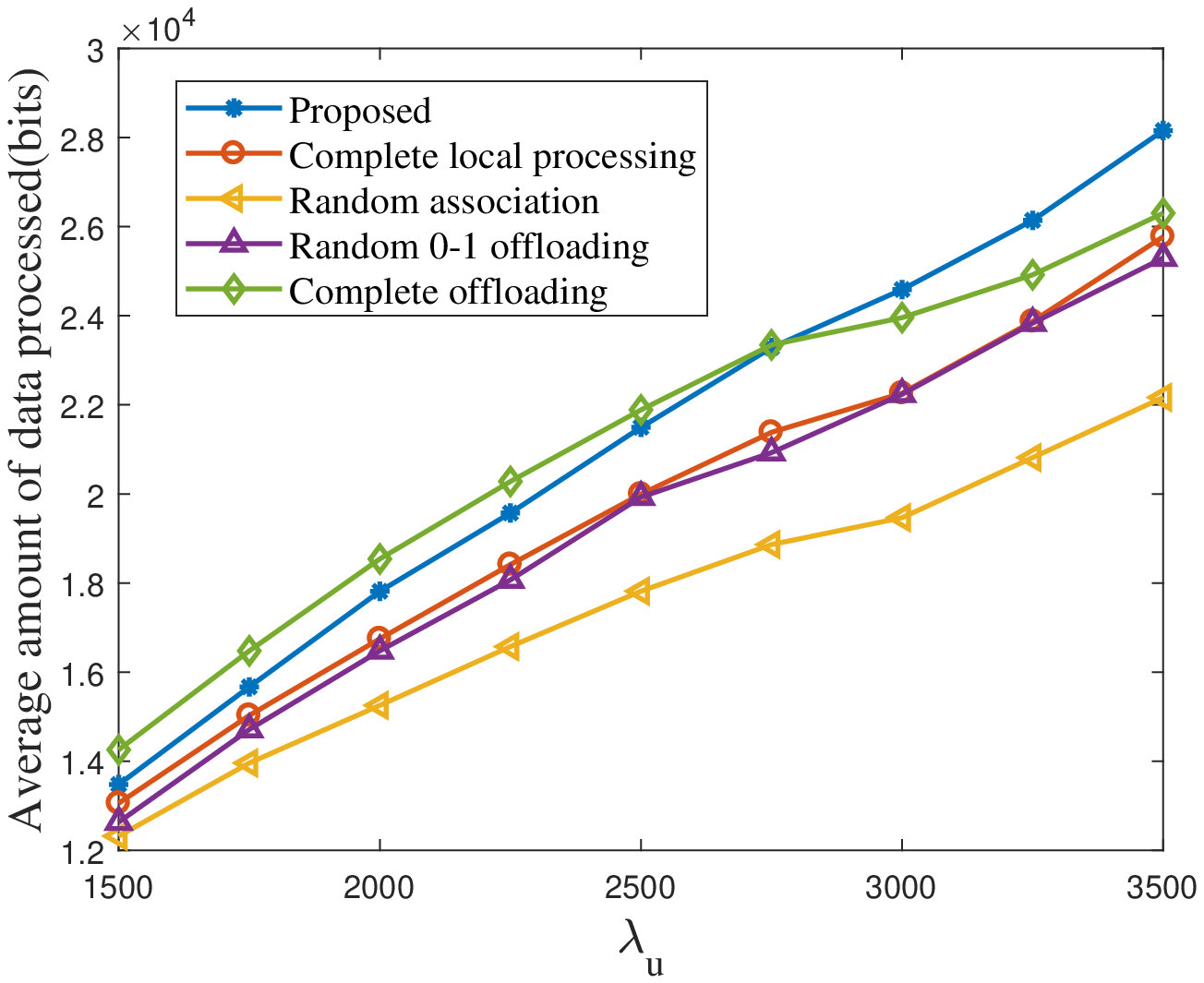}	
		\caption{Processing data v.s. average task size $\lambda_u$.\label{lambda-pro}}
	\end{minipage}
\end{figure}

In Figs. \ref{lambda-ee}-\ref{lambda-pro}, we evaluate the impact of the average task arrival size  $\lambda_u$ on the system performance for all different  schemes. 
Fig. \ref{lambda-ee} illustrates the relationship between the network EE and the average task size. The proposed algorithm achieves the best network EE compared with the other four  benchmark algorithms and it further verifies the efficiency of the proposed method. As the average task size $\lambda_u$  increases, the network EE of all the schemes deteriorates as each MID needs to consume more energy to increase the data processing rate (which can be further observed in Fig. \ref{lambda-ener}). As the  proposed algorithm can process tasks parallelly and allows each MID to select the best MEC server, its energy consumption is lowest and the processing data is almost the highest (slightly lower than the `complete offloading' method when task size is at a lower level which can be observed from Fig. 11), resulting in the smallest network EE.
On the other hand, the random association scheme has the worst network EE. This is due to the fact that random MEC association leads to a poor offloading rate, thus degrading the EE. The complete local method has the second-worst EE due to the limitation of the local computational resources.

Fig. \ref{lambda-delay} shows the service delay versus the average task size. 
As the task arrival size $\lambda_u$ increases, the average queue length increases accordingly, which leads to an increase in the service delay.
Note that when the task size is relatively small, the service delay under the `complete offloading' algorithm is lower than the proposed method whereas it is just the opposite when the task size is at a large level. The explanation is given as follows.  The `complete offloading' scheme cannot benefit from the local processing. {When the channel condition is poor, the task queue length keeps accumulating but will maintain a low value because of the relatively small task arrival size. Once the channel condition becomes better, the improved instant processing rate can rapidly reduce the queue backlog, finally lead to a lower average delay. }Therefore, the average delay is better than the proposed algorithm in this circumstance.
On the other hand, the larger task arrival size results in a large backlog when the channel condition is not good, so the instant improved processing data when the channel becomes better cannot rapidly reduce the average queue backlog. 
But the queue length under our proposed algorithm can maintain at a stable level because of the partial processing capability, which leads a better delay performance when the task arrival size is at a large level.

The average energy consumption versus $\lambda_u$ is shown in Fig. 10.
It can be seen that the energy consumption increases with the average task arrival size. As the principle of the Lyapunov optimization algorithm is to optimize the objective but at the same time to keep the queue stable. When the task size increases, the system needs to increase the processing rate to ensure the stability of the queue and the user's QoS, resulting in an increase in energy consumption and a decrease in EE, which is consistent with the phenomenon observed in the previous figures. In addition, the proposed method can take advantage of the local processing, remote processing, and the dynamic MEC association, which helps to save as much energy as possible. Therefore, the average energy consumption of the proposed method stays at the lowest than other benchmarks. Furthermore, the benchmarks with MEC processing, i.e., `Complete offloading' and `Random 0-1 offloading' schemes,  can achieve a lower energy consumption than the benchmark without MEC processing, i.e. `Complete local processing', because of the powerful executing capability of the MEC server which can enable the MIDs to complete their task computation more efficiently. Thus, the benefits of deploying MEC servers have been verified.

The average processing data versus $\lambda_u$ is presented in Fig. 11, the processing data increases with the average task arrival size. This trend is aligned with our observation in the previous plots. One thing is worth noting that although the processing rate of the proposed algorithm is lower than that of the complete offloading scheme when the task size is small, our proposed method aims to save more energy to obtain the optimal network EE, leading to a lower processing rate. Another phenomenon we can observe is that the 'Random association' method has the lowest average amount of data processed. The reason is that if the association indicator is not optimized, a mal-performing MEC  can be selected for offloading, which causes waste of energy and the low system processing rate. This observation also verifies the merit of our proposed algorithm.

\section{Conclusions}
In this paper, we investigate the fundamental EE-delay tradeoff in multi-user multi-server IoT networks, jointly taking stochastic task generating, wireless channel environment, and user mobility into consideration. The long-term average network energy efficiency minimization problem subject to available resources constraint while maintaining queuing stability was formulated. To solve this problem, we proposed an online  offloading and resource allocation algorithm based on Lyapunov optimization and the submodular method for mobile IoT networks. Theoretical analysis indicates that the network EE and service delay obey an $[O(1/V), O(V)]$ tradeoff with $V$ as a control parameter. Extensive simulation validates the tradeoff and demonstrates our proposed scheme outperforms other benchmark schemes in terms of the EE-delay performance. {In our future work, we will expand our research by adopting new techniques such as machine learning and considering the privacy-preserving \cite{new2-privacy} in the multi-server multi-user IoT networks.}

\section*{APPENDIX}
\setlength{\abovecaptionskip}{-0.2cm} 
\setlength{\belowcaptionskip}{-1cm}
\subsection{Proof for Lemma 2}
\setlength{\abovecaptionskip}{-0.2cm} 
\setlength{\belowcaptionskip}{-1cm}
By squaring both sides of equation (4), we have
\be
	\setlength{\abovedisplayskip}{3pt}
\setlength{\belowdisplayskip}{3pt}
\begin{aligned}
	Q_u^l&(t+1)^2=(Q_u^l(t)-D_u^l(t))^2+(c_u(t)A_u(t))^2\\ 
	&+2c_u(t)A_u(t)[Q_u^l(t)-D_u^l(t)]^+\\  
	&\le (Q_u^l(t)-D_u^l(t))^2+(c_u(t)A_u(t))^2+2c_u(t)A_u(t)Q_u^l(t)\\   
	&=Q_u^l(t)^2-2Q_u^l(t)[D_u^l(t)-c_u(t)A_u(t)]\\ 
	&+A_u(t)^2c_u(t)^2+D_u^l(t)^2.
\end{aligned} 
\ee
Summing over all $u (u=1,...U)$ of above inequality, the following inequality can be obtained:
\be
	\setlength{\abovedisplayskip}{3pt}
\setlength{\belowdisplayskip}{3pt}
\begin{aligned}
	\frac{1}{2}\sum\limits_{u = 1}^U [Q_u^l(t+1)^2-&Q_u^l(t)^2]\le \sum\limits_{u = 1}^U {{A^2_u(t)c^2_u(t)+D_u^l(t)^2}\over{2}}\\
	&-\sum\limits_{u = 1}^U[Q_u^l(t)(D_u^l(t)-c_u(t)A_u(t))].  \label{11}
\end{aligned}
\ee
Similar manipulations can be used in (5),
\be
\begin{aligned}
	\frac{1}{2}\sum\limits_{u = 1}^U [Q_u^o(t+1)^2-&Q_u^o(t)^2]\le \sum\limits_{u = 1}^U {{A^2_u(t)(1-c_u(t))^2+D_u^o(t)^2}\over{2}}\\
	&-\sum\limits_{u = 1}^U[Q_u^o(t)(D_u^o(t)-(1-c_u(t))A_u(t))].\\ \label{22}
\end{aligned}
\ee
Summing up (\ref{11}) and (\ref{22}) yield
\be
\begin{aligned}
	L(&\mathbf{\Theta} (t + 1)) - L(\mathbf{\Theta} (t))\le -\sum\limits_{u = 1}^U[Q_u^l(t)(D_u^l(t)-c_u(t)A_u(t))]\\
	&-\sum\limits_{u = 1}^U[Q_u^o(t)(D_u^o(t)-(1-c_u(t))A_u(t))]\\
	&+\sum\limits_{u = 1}^U {{A^2_u(t)[c^2_u(t)+(1-c_u(t))^2]+D_u^l(t)^2+D_u^o(t)^2}\over{2}}. \label{33}
\end{aligned}
\ee
Let  $C_1=\frac{1}{2}\sum_{u=1}^U[(D^{l}_{u,\max})^2+(D^{o}_{u,\max})^2]
	\geq \frac{1}{2}\sum\limits_{u = 1}^U {[D_u^l{{(t)}^2} + D_u^o{{(t)}^2}} ]$, $D^{l}_{u,\max}$ and $D^{o}_{u,\max}$ are the upper bound of $D_u^l(t)$ and $D_u^o(t)$, respectively. Substituting $C_1$ and (\ref{33}) into the right side of (\ref{drift}), we can obtain the desired result.

\subsection{Proof for Theorem 1}
 Given a ground set $\mathcal{G}(t)$ and a subset $\chi (t) \in \mathcal{G}$, the set function $J(\chi)$ is maximizing monotone submodular set function.
 
(1) if $\chi=\emptyset$, $J(\emptyset ) = 0$.

(2) Given two arbitrary subsets $X$ and $Y$ $(Y(t) \subseteq X(t) \subseteq \chi (t))$, and an arbitrary element ${\tilde{x}_{um}}(t) \subseteq G(t)\backslash X(t)$, after adding ${\tilde{x}_{um}}(t)$ into $X$ and $Y$, we have the following relationship:
\be
	\begin{aligned}
	J(Y(t)& \cup {\tilde{x}_{um}}(t)) - J(Y(t))\\
	&= J(X(t) \cup {\tilde{x}_{um}}(t)) - J(X(t)) = {W_{um}(t)}.
	\end{aligned}
\ee
Therefore, we can deduce that $J(\chi (t))$ is a monotone submodular function and the  $\text{P}_{3.4.5}$ can be formulated as the maximization of a monotone submodular problem with two partition matroid constraints.

\subsection{Proof for Theorem 2}
Based on Lemma 2, we can obtain the following inequation:
\be
	\setlength{\abovedisplayskip}{3pt}
\setlength{\belowdisplayskip}{3pt}
\begin{array}{l}
	\Delta (\mathbf{\Theta} (t)) + V \mathbb{E}\left\{E(t)-\eta_{EE}(t) \sum_{u=1}^U\{D_u^{l}(t)+D_u^{o}(t)\}\right\}\\
	\le C_1+V\mathbb{E}\left\{E^{\Pi}(t)-\eta_{EE}(t) \sum\limits_{u = 1}^U [ D_u^{l,\Pi}(t)+D_u^{o,\Pi}(t)] \right\}\\
	- \mathbb{E}\left\{\sum\limits_{u = 1}^U Q_u^l(t)\left[D_u^{l,\Pi}(t)-c_u^\Pi(t)A_u(t)\right] \right\}\\
	- \mathbb{E}\left\{\sum\limits_{u = 1}^U Q_u^l(t)\left[D_u^{o,\Pi}(t)-(1-c_u^\Pi(t))A_u(t)\right] \right\}\\
	+A^2_u(t)\left[(c_u^\Pi(t))^2-c_u^\Pi(t)+\frac{1}{2}\right],
	\\ \label{driftpenalty1}
\end{array}
\ee
where $E^{\Pi}(t)$, $D_u^{l,\Pi}(t)$, $D_u^{o,\Pi}(t)$ and $c^\Pi _u(t)$ are the resulting values under the independent, stationary, and randomized algorithm.
Plugging  (\ref{eq1})-(\ref{eq3}) into (\ref{driftpenalty1}), 
\be
	\setlength{\abovedisplayskip}{3pt}
\setlength{\belowdisplayskip}{3pt}
\begin{array}{l}
	\Delta (\mathbf{\Theta} (t)) + V \mathbb{E}\left\{E(t)-\eta_{EE}(t) \sum_{u=1}^U\{D_u^{l}(t)+D_u^{o}(t)\}\right\}\\
	\le C_1+C_2+V\mathbb{E}\Big{\{}(\eta^*_{EE}+\delta-\eta_{EE}(t)) \sum\limits_{u = 1}^U [ D_u^{l,\Pi}(t)\\
	+D_u^{o,\Pi}(t)] \Big{\}}-\epsilon \sum\limits_{u = 1}^U [ Q_u^{l}(t)+Q_u^{o}(t)],  \label{driftpenalty2}
\end{array}
\ee
where $ C_2=\frac{A^2_{max}}{2}$.
Taking a limit as $\delta \rightarrow 0$, and summing up both sides of $(\ref{driftpenalty2})$ over $t \in \{0,1,...T-1\}$  yield
\be
	\setlength{\abovedisplayskip}{3pt}
\setlength{\belowdisplayskip}{3pt}
\begin{array}{l}
	\mathbb{E} \{L(\Theta (T))\} -\mathbb{E}\{L(\Theta (0))\} \\
	+ V \sum\limits_{t=0}^{T-1} \mathbb{E}\left\{E(t)-\eta_{EE}(t) \sum\limits_{u=1}^U\{D_u^{l}(t)+D_u^{o}(t)\}|\mathbf{\Theta} (t)\right\}\\  
	\le (C_1+C_2)T+V\sum\limits_{t=0}^{T-1}\left[  \eta_{EE}^*-\mathbb{E}\{\eta_{EE}(t)\}    \right]
	(D^{l}_{\min}+D^{o}_{\min})\\-\epsilon \sum\limits_{t=0}^{T-1}\sum\limits_{u = 1}^U [ Q_u^{l}(t)+Q_u^{o}(t)],  \label{driftpenalty3}
\end{array}
\ee
where $ D^{l}_{\min}=\min(\{\sum\limits_{u=1}^{U}D_u^l(t)\})$ and $ D^{o}_{\min}=\min(\{\sum\limits_{u=1}^{U}D_u^o(t)\})$.
Since $ Q_u^{l}(t) \ge 0$, $ Q_u^{o}(t) \ge 0$, and $\mathbb{E} \{L(\Theta (T))\}\ge 0$,  divided by $VT$ in each side, after neglecting the last term, rearranging (\ref{driftpenalty3}), we have
\be
	\setlength{\abovedisplayskip}{3pt}
\setlength{\belowdisplayskip}{3pt}
\begin{array}{l}
	\frac{1}{T}\sum\limits_{t=0}^{T-1}\left[\eta_{EE}^*-\mathbb{E}\{\eta_{EE}(t)\}\right](D^{l}_{\min}+D^{o}_{\min })\\
	-\frac{1}{T}\sum\limits_{t=0}^{T-1} \mathbb{E}\left\{E(t)-\eta_{EE}(t) \sum\limits_{u=1}^U\{D_u^{l}(t)+D_u^{o}(t)\}|\mathbf{\Theta} (t)\right\}\\ 
	+\frac{1}{T}\mathbb{E}\{L(\Theta (0))\}+\frac{(C_1+C_2)}{V} \ge 0.  \label{driftpenalty4}
	
\end{array}
\ee
Let $T$ go to infinity, we get the following inequality:
\be
	\setlength{\abovedisplayskip}{3pt}
\setlength{\belowdisplayskip}{3pt}
\begin{array}{l}
	(\eta_{EE}^*-\eta_{EE})(D^{l}_{\min}+D^{o}_{\min })+\frac{C_1+C_2}{V}
	\ge 0.	                  \label{driftpenalty5}
\end{array}
\ee
Rearranging (\ref{driftpenalty5}) yields the upper bound of the network EE under our proposed algorithm.
Similarly, dividing (\ref{driftpenalty3}) by $\epsilon T$ and taking a limit as $T \rightarrow 0$,  we obtain
\be
	\setlength{\abovedisplayskip}{3pt}
\setlength{\belowdisplayskip}{3pt}
\begin{array}{l}
	\overline Q_{\Sigma}=\sum\limits_{T \to +\infty } \frac{1}{T} \sum\limits_{t = 0}^{T - 1}\sum\limits_{u = 1}^{U} \mathbb{E}\{Q^l_u(t)+Q^o_u(t)\}\\
	\le {    {   C_1+C_2+V\eta_{EE}^*\sum\limits_{u=1}^U\{D_u^{l}(t)+D_u^{o}(t)\}-{{\eta _{EE}}(t)\sum\limits_{u = 1}^U {[D_u^{l,\Pi }(t) + D_u^{o,\Pi }(t)]} }
		 } \over   {\epsilon }     }\\
	\le { {   C_1+C_2+V\eta_{EE}^*(D^{l}_{\max}+D^{o}_{\max })-E^{min} } \over {\epsilon }},
\end{array}
\ee
where $ D^{l}_{\max}=\max(\{\sum\limits_{u=1}^{U}D_u^l(t)\})$, $ D^{o}_{\max}=\max(\{\sum\limits_{u=1}^{U}D_u^o(t)\})$, and $E^{min}$ is the minimum network energy consumption.

\end{document}